\documentclass[10pt,preprint]{aastex}

\def\Fe{iron K$\alpha$~}
\def\exp{\rm e}

\def\ltsima{$\; \buildrel < \over \sim \;$}
\def\simlt{\lower.45ex\hbox{\ltsima}}
\def\gtsima{$\; \buildrel > \over \sim \;$}
\def\simgt{\lower.45ex\hbox{\gtsima}}

\received{April 12, 2001}
\accepted{July 12, 2001}

\shorttitle{relativistic bulk motion in corona and the \Fe line}
\shortauthors{Lu \& Yu}

\begin{document}
\title{The effects of relativistic bulk motion of X-ray flares in
the corona on the iron K$\alpha$ line in Seyfert 1 galaxies}
\author{Youjun Lu\altaffilmark{1}
\affil{Center for Astrophysics, Univ. of Sci. \&
Tech. of China, Hefei Anhui 230026, P.\ R.\ China \\
National Astronomical Observatories, Chinese Academy of Sciences}
       Qingjuan Yu
\affil{Princeton University Observatory, Peyton Hall, Princeton, NJ
08544-1001, USA}}
\email{lyj@astro.princeton.edu; yqj@astro.princeton.edu}
\altaffiltext{1}{Present Address: Princeton University Observatory, 
Peyton Hall, Princeton, NJ 08544-1001, USA}

\begin{abstract}

It is likely that dramatic magnetic flares in a corona above a black hole
accretion disk dominate the X-ray emission in Seyfert 1 galaxies. 
Such flares are likely to move with mildly relativistic
bulk velocity. We study the effects of the bulk motion of X-ray flares on
the shape and equivalent width of the \Fe line from an untruncated cold 
disk around a Kerr black hole using fully relativistic calculations. The 
flares are located above a cold accretion disk$-$either on or off the 
rotation axis. The upward/outward bulk motion of flares causes a reduction
of the \Fe line width, while the downward/inward bulk motion of flares 
causes an increase of the \Fe line width.
To a distant observer with a low 
inclination angle ($\theta_{\rm o} \simlt 30\arcdeg$), larger upward/outward
bulk velocities decrease the extension of the red wing, with little change
in the location of the blue `edge'. In contrast, an observer at a large 
inclination angle (e.g. $\theta_{\rm o}=60\arcdeg$) sees both the red wing
and the blue `edge' change with the bulk velocity. The equivalent width of
the \Fe line decreases rapidly with increasing bulk velocity of flares. 
However, the `narrower' line profiles observed in some objects (e.g. IC4329A
and NGC4593) are difficult to produce using the out-flowing magnetic flare
model with an appropriate equivalent width unless the X-ray emission is
concentrated in an outer region with a radius of several tens of $r_{\rm g}
=GM/c^2$ or more. This suggests that other parameters, such as the 
`truncation radius' due to disruption or ionization of the inner disk may
still be needed. An important result is that the \Fe line intensity is found
to be constant even though the continuum flux varies significantly, which
is true for out-flowing magnetic flares with different bulk velocities but
similar intrinsic luminosities when located close to the central black
hole. This is caused by the combination of the effects of relativistic 
beaming and gravitational lensing. We find that fluctuations in the bulk
velocities of out-flowing low-height flares located at the inner region
($r\simlt 15r_{\rm g}$) can account for a constant \Fe line and significant
continuum variation as observered in MCG-6-30-15 and NGC5548. This is 
especially interesting for MCG-6-30-15 whose behavior is difficult
to explain using the ionization model.

\end{abstract}


\section{Introduction} \label{sec:intro}

Two prominent X-ray features, the \Fe line at 6.4~keV and the broad
Compton reflection hump, have been found in Active Galactic Nuclei (AGNs)
and in galactic black hole candidates (GBHCs) using observations made with
the {\it Ginga, ASCA, RXTE}, and {\it BeppoSAX} satellites.
These features have been explained as the Compton re-processing
imprints of the hard X-ray power-law continuum impinging on the surface
layer of a thin accretion disk \citep[and references therein]{gf91,mpp91}.
The \Fe line is thought to be produced by the fluorescence of the cold
disk material and is accompanied by the Compton-scattered continuum.

The \Fe line was first observed by {\it Ginga} (e.g. Pounds et al. 1990;
Nandra \& Pounds 1994).  
The observations from {\it ASCA} have since confirmed that the line profile
is both very broad and skewed \citep[]{tan95,nan97}; these properties are 
generally believed to be caused by both special relativistic (i.e. 
transverse-Doppler and special relativistic boosting) and general 
relativistic (i.e. gravitational redshifts and light bending) effects. 
\citep[]{fab89,lao91,koj91}. In MCG-6-30-15, the line profiles provide 
evidence of line emission from the region very close to the central black 
hole, at a radius less than $5r_{\rm g}$ (where $r_{\rm g}=GM/c^2$ is the
gravitational radius, $G$ is the gravitational constant; and $c$ is the speed
of light; hereafter $G=c=1$ and thus $r_{\rm g}=M$) \citep[]{iwa96,iwa99}. 
So far, alternative models have failed to consistently interpret the broad 
\Fe line profile and the continuum spectrum \citep[]{fab95,fab00}. 

There are two main aspects of the theoretical work on the broad \Fe line.
One aspect is to study the detailed properties of the ``reflector'', i.e. the 
accretion disk. Considering the geometry, size and ionization state of the
accretion disk, reflection spectra have been computed by performing 
simulations of Compton reflection from neutral or ionized 
matter, including fluorescent iron emission 
\citep[]{gf91,mpp91,mat92,zc94,mfr93,mfr96,rfb96,rfy99,nkk00}. 
The finite thickness, radial accretion flow, turbulence \citep[]{pb98} and
warp of the disk \citep[]{hb00} have also been considered and shown to 
significantly affect the line emission. The second aspect is the study of 
the effects of primary X-ray emission sources (which provide energetic 
X-ray photons to illuminate the accretion disk) on the line properties.
Dramatic magnetic flares probably dominate the X-ray emission in AGNs, and
the general axisymmetric emissivity law is therefore not realistic for a
time-dependent study due to the arbitrary location of X-ray flares (though
it may be appropriate for some time-averaged studies). Some simple models 
have shown that the line profile and equivalent width (EW) change significantly 
with the location of the flares \citep[]{rey99,yl00,rus00,dl01}. In this
paper, we follow the second aspect and study the effects of the bulk motion
of X-ray emitting flares on the \Fe line properties.

The reflection fraction $R=\Omega/2\pi$, where $\Omega$ is the solid angle
subtended by the reflecting material, was found to be in the range $0-1$ or
even $>1$ in some objects observed by {\it Ginga} and {\it BeppoSAX}
\citep[]{pounds90,np94,mat00}. The average reflection fraction is around
$0.5$. An interesting correlation between the reflection fraction $R$ and 
the photon index $\Gamma$ for both AGNs and GBHCs has been claimed by 
\citet[]{zls99}. On average, sources with flat (i.e. smaller photon index)
continuum spectra tend to have
a low reflection fraction. This relation is interpreted as follows: The cold
(neutral or weakly ionized) thin disk extends down to only several tens of
$M$, and within this radius is either disrupted \citep[]{emn97,pkr97} 
or totally ionized 
\citep[]{rfy99,nkk00}; this leads to a reflection fraction in
the range $0-1$. An alternative model is that the reduction (or enhancement)
of the reflection fraction is caused by mildly relativistic outflow (or 
inflow) in the corona above an untruncated accretion disk \citep[]{bel99b}. 
In this model, the reflection fraction can be either $>1$ or $<1$, and a good
fit of the $R-\Gamma$ relation can be obtained \citep[]{bel99b,mbp01}. 
However, one
should be cautious about the $R-\Gamma$ relation since recent investigations
have shown that this correlation is at least partly caused by systematic
errors in data fitting \citep[]{nan00,ve01}. The extensive simulations
of \citet[]{ve01} show that the correlation can result from measuring
the two quantities from the same spectra. Further investigations are needed
to determine whether a real correlation between $R$ and $\Gamma$ exists.

The Monte-Carlo simulations of \citet[]{gf91} have shown a relation between
the reflection fraction $R$ and the EW of \Fe lines [EW(K$\alpha$) $\sim 130
R$~eV for a typical photon index of Seyfert 1 galaxies with $\Gamma=1.90$
and for solar metallicity]. 
For a fixed disk/corona geometry and ionization, i.e. where $R$ is a constant,
the \Fe line EW is expected to remain constant and the line flux increases
with increasing continuum flux if the probing timescale is longer than the 
light crossing time of the emission-line region, which is generally true for
{\it ASCA} observations of AGNs. If there is a change in the geometry of the
system, the line EW is expected to vary in step with the reflection fraction
$R$. From observations using {\it ASCA}, the average EW of \Fe lines for 
Seyfert 1 galaxies is about $200-300$~eV \citep[]{nan97}, which is considerably
larger than expected from the $R-$EW relation and the observed $R$ value ($R$
typically $<1$). It is difficult to understand this difference since the 
fluorescence must be accompanied by the Compton-scattered continuum. Recently,
\citet[]{lz01} re-analyzed the average spectra and obtained average line 
profiles for three sets of Seyfert 1 galaxies grouped according to the 
spectral index of the X-ray continuum. Modeling the profile with a disk line 
plus a narrow line component, they found that the EW is consistent with $R$
following the expected relation above. However, it is still puzzling that in
some bright objects the line flux appears constant (or the EW is inversely 
proportional to the continuum flux) and the EW is anti-correlated with $R$ 
while the continuum flux varies strongly. For example, the relationship
between the EW and the continuum flux ($F$) in NGC5548 can be well fit by a 
power-law EW(K$\alpha$)$\propto F^{\alpha}$ with a slope of $\alpha=-0.9
\pm0.4$; this is consistent with a constant line flux (see Fig.~4 in Chiang 
et al. 2000). Similar results are observed in MCG-6-30-15 (see Table~4 and 
Fig.~13b in Lee et al. 2000). The reason that $R$ fails to trace the variation
in the EW is still unknown. Two possible mechanisms, the thermal instability 
of photon-ionized plasma leading to a highly ionized skin on the surface of
the thin accretion disk \citep{nkk00} and resonant trapping followed 
by Auger destruction \citep[]{rf93} may suppress the \Fe line emission
relative to the reflection continuum \citep[]{reynolds00,brf01,zr01}.

The rapid X-ray variability in many Seyfert 1 galaxies suggests that the X-ray
is likely to be emitted during dramatic flare-like events (e.g. the light
curve of MCG-6-30-15 in Iwasawa et al. 1999). Magnetic flares have long been
considered as the source of rapid variability for accreting black holes
\citep[]{grv79}. The variability of the X-ray source can induce the
corresponding variation in the \Fe line. \citet[]{fab89} suggested that any
variation of the primary X-ray source would be `echoed' by the evolution of
the \Fe line profile, which was then considered in detail by \citet[]{stella90},
\citet[]{mp92} and \citet[]{cs93,cs95}, the latter for different source
geometries. Recently, \citet[]{rey99} generalized this work, assuming an
instantaneous static X-ray flash (on- or off-axis) atop a thin
disk. Their calculations of \Fe line reverberation show that it is
possible to probe both the mass and spin of black holes in AGNs from the
reverberation mapping signatures using the planned X-ray satellite {\it
Constellation X-ray Mission} \citep[]{yr00}.

It is likely that the dramatic magnetic flares in the inner region 
(e.g. $r\simlt 15 M$) dominate
the X-ray emission \citep[]{mf01} and move with mildly relativistic bulk 
velocity \citep[]{rf97,bel99a}; this may play an important role in producing
the reflection and the fluorescent iron line.  Reynolds \& Fabian (1997) 
estimated the special relativistic effects of the bulk
motion of the X-ray emitting material (relative to the disk) on the line EW.
They found that the line EW is significantly enhanced (or reduced) if the 
bulk motion of the X-ray source is moving directly towards (or away from) 
the disk, as intuitively expected. The line EW is also found to be enhanced
(but to a lesser degree) if the X-ray emitting material moves along the surface
of the disk.  The X-ray reflection from the disk is also reduced (or enhanced)
due to bulk motion of the X-ray source away from (towards) the reflector, 
i.e., disk, by the special relativistic beaming effect \citep[]{bel99b}. 
For flares with bulk motion, the corresponding reverberation mapping
properties are also expected to differ from those derived for static flares
by \citet[]{rey99}. In the present paper, we perform a complete relativistic
calculation (combining both special relativistic and general relativistic
effects simultaneously) to investigate the impact of the bulk motion of flares
in the corona on the profile and EW of the \Fe line (the ionization of
the reflecting material is not considered). We describe the model in \S2. 
Calculation results are presented in \S3, The discussion and applications
are presented in \S~4 and conclusions are given in \S5.

\section{Model assumptions and calculations} \label{sec:model}

\subsection{The X-ray flare and the disk illumination} \label{sec:mod_ill}

We shall assume that the primary X-ray flares are located above the inner
region of an accretion disk. 
In the following calculation, we assume the flare is not instantaneous
flash but a constant source of illumination (i.e. not time dependent).
Assuming that the flare is point-like, the X-ray photons emitted by the flare
can freely propagate to the observer, the accretion disk and the black hole 
without absorption or attenuation.  If the flaring plasma in the corona is 
composed of electron-positron ($e^{\pm}$) pairs, it should
be accelerated away from the disk by the pressure of soft radiation which
dominates the bolometric luminosity of the disk \citep[]{bel99a}. The bulk
velocity is expected to be in the range 0.1---0.7 \citep[]{bel99b}.
\citet[]{fier93} argued that magnetic instabilities and reconnection
events in a disk/corona could produce shock waves and/or streaming of
relativistic particles along the magnetic field lines, similar to the 
ejection in solar flares; the corresponding X-ray emitting plasma may
therefore be in bulk motion relative
to the disk. The plasma can be ejected away or towards the disk,
with the preferential direction being away from the disk \citep[]{bel99c}.
It is then reasonable to assume that the flare moves with mildly 
relativistic bulk velocity upward (or downward) relative to the accretion 
disk or outward (or inward) along the radius due to the radiative acceleration 
\citep[]{bel99a} or ejection \citep[]{fier93}. In order to trace the 
photon trajectories, we first derive the formulae that express
the constants of motion for photons originating from a flare with arbitrary
bulk motion in the vicinity of a black hole (using Boyer-Lindquist coordinates).
These are listed in Appendix A. The Monte-Carlo simulations of \citet[]{jcz00}
show that the anisotropy effect in the hard X-ray emission from corona
with bulk motion, which is dominated by multiple-Compton-scattering photons,
is weak. Thus it is reasonable to adopt the X-ray emission from a
flare with bulk motion as an isotropic distribution in the rest frame of the
flare. We model the isotropic distribution of radiation in the rest frame of
the X-ray flare by a Monte-Carlo method, keeping track of the Boyer-Lindquist 
coordinates of each photon from the given source position ($r_{\rm s}$, 
$\theta_{\rm s}$, $\varphi_{\rm s}$) to position on the disk ($r_{\rm d}$, 
$\pi/2$, $\varphi_{\rm d}$) using a ray tracing technique and elliptic 
integrals (e.g. Rauch \& Blandford 1994), and thus we can easily obtain the 
illumination flux grid as a function of time and energy in the rest frame of
the co-rotating disk material; we account for both special relativistic 
(Doppler shift and boosting) and general relativistic (gravitational redshift
and gravitational lensing) effects (see Appendix B and \S~\ref{sec:mod_proew}).
We also adopt the velocity field of the accretion disk given by \citet[]{cun75}.
The formulae for calculating the illumination flux are listed in Appendix B. 
The associated redshift factor and the travel time from the source to the 
accretion disk have also been calculated.

The X-ray photons impinging onto the disk plane will be reprocessed and
reflected. The reflected spectrum and associated iron fluorescent
line depend on the disk structure and the ionization state of the disk
material. We assume the disk is geometrically thin and optically thick.
The observed \Fe line profiles in some Seyfert 1 galaxies (e.g.
MCG-6-30-15) suggest the existence of cold material in the region within 
a radius of several $M$. We assume the disk material is relatively cold and
sufficiently dense so that the ionization can be neglected in the
region outside the marginal stable orbit $r_{\rm ms}$. However, the detailed
nature of the accretion flow within $r_{\rm ms}$ is very complicated.
\citet[]{rb97} pointed out that the fluorescent line photons can be
emitted from the region within $r_{\rm ms}=6M$ for a Schwarzschild
black hole since the accretion flow does not immediately become optically
thin. However, the continuum spectrum expected from this model will have
a large absorption edge associated with it, which may be absent in
the {\it ASCA} data \citep[]{yrf98}.
There are different views about the dynamics at the inner edge of the 
accretion disk: the presence of magnetic fields may exert a torque on the
inner edge of the disk in contrast to the usual assumption of a zero-torque
boundary condition (Agol \& Krolik 2000; but Armitage, Reynolds \& Chiang
2001). It is therefore difficult to calculate with this region. In fact, this
region is only relevant to the iron line emission in the case of a 
Schwarzschild black hole since the marginal stable orbit of maximum Kerr 
black hole ($a/M=0.998$) approaches the event horizon. The observations 
suggest the existence of Kerr black holes in objects like MCG-6-30-15 and
MRK766 \citep[]{dab97,bra01}. Thus we consider the case of a maximum Kerr
black hole ($a/M=0.998$) in the present paper; a Schwarzschild black hole
will be considered in a future study.

\subsection{The iron K$\alpha$ line profile and equivalent width } 
\label{sec:mod_proew}

To calculate both the time-average or time-varying iron line
profiles and corresponding EW, we use the following prescription:

1) We use a ray-tracing technique and elliptic integrals \citep[]{rb94}
to follow the trajectories of photons from the observer, keeping track of
all Boyer-Lindquist coordinates $t$, $r$, $\theta$ and $\varphi$
until the photons either intersect the accretion disk plane, disappear
below the event horizon or escape to `infinity' (operationally defined
to be $r=1000M$). (Note here that the spin parameter should be changed
from $a$ to $-a$ since we use time-reversed photons to trace the tracks of
out-coming photons.) 
We then calculate the redshift factor for a photon (to the observer)
emitted from a particular position on the disk (see Appendix B). The solid
angle subtended at the observer by each disk grid element is also calculated.
We set the inner radius to be at the marginally stable orbit (only few \Fe 
photons come from the region within the $r_{\rm ms}$ for Kerr
black hole with spin $a/M=0.998$), and the outer radius at $160M$
(setting a larger outer radius will not significantly affect the line 
properties since most of the \Fe line photons are emitted in the inner disk
region).

2) The illumination law of a flare can be derived by Monte-Carlo simulation
(see Appendix B and \S~\ref{sec:mod_ill}). The fluorescent process was
described in detail by \citet[]{gf91}.  Following them, we define
$N_{E_{\rm d}}^{\rm in}(E_{\rm d},\theta_{\rm in}; r, \varphi)$ to be the
number flux of photons impinging onto the disk grid element ($r$, $\varphi$)
with energy $E_{\rm d}$ and incident angle $\theta_{\rm in}$ per unit 
frequency and per unit time, as measured by an observer co-rotating with the
disk (see the formula in Appendix B). The incident angle $\theta_{\rm in}$
is determined by $\theta_{\rm in}=\cos^{-1}(-{\bf p}\cdot{\bf n}/{\bf p}
\cdot {\bf u}_{\rm d})$, where ${\bf p}$ is the photon's momentum, ${\bf n}$
is the surface normal of the disk and ${\bf u}_{\rm d}$ is the 4-velocity 
of the accretion disk material (see also Appendix B). Thus
the number flux of fluorescent photons per unit frequency and per unit time
which are able to escape the disk is
\begin{equation}
N^{\rm out}_{E_{\rm d}}=N^{\rm in}_{E_{\rm d}}(E_{\rm d,} \theta_{\rm in}; r, \varphi)
Y(\theta_{\rm in},E_{\rm d}),
\end{equation}
where the yield $Y(\theta_{\rm in},E_d)$ can be expressed as: 
\begin{equation}
Y(\theta_{\rm in},E_{\rm d})=\eta(\theta_{\rm in})f(E_{\rm d}),
\end{equation}
and 
\begin{equation}
\eta(\theta_{\rm in})=(6.5-5.6\cos\theta_{\rm in}+2.2\cos^2\theta_{\rm in})\times 10^{-2},
\end{equation}
\begin{equation}
f(E_{\rm d})=7.4\times10^{-2}+2.5{\rm exp}(-\frac{E_d-1.8}{5.7}).
\end{equation}
This analytical approximation is valid for $E_{\rm t}<E_{\rm d}<E_{\rm m}$,
where $E_{\rm t}=7.1$~keV
is the energy threshold for triggering the 6.4~keV fluorescent line, and 
$E_{\rm m}=30$~keV. We also include the anisotropic effect of the emerging
fluorescent emission through the simple assumption that the line emission is 
proportional to $\eta'(\theta_{\rm out})=2\cos\theta_{\rm out}\ln
(1+1/\cos\theta_{\rm out})$ \citep[]{bas78,haardt93,ghm94}, where
$\theta_{\rm out}$ is the outgoing inclination angle measured in the
co-rotating frame, which is determined by the same formula as
$\theta_{\rm in}$.

The contribution of each grid element on the accretion disk surface to
the iron line is given by the transfer function corresponding to a
given position of the flare,
\begin{eqnarray}
\Psi(E)=\frac{1}{\Delta E}
\sum_{
E\rightarrow E+\Delta E
}
g_{\rm do}^4 \eta'(\theta_{\rm out})\frac{\delta\Omega_{\rm do}}{4\pi} \nonumber \\
\cdot
\delta(E-E_{\rm Fe}/g_{\rm do}) E_{\rm Fe}\int^{E_{\rm m}}_{E_t}N^{\rm out}_{E_{\rm d}}dE_{\rm d},
\end{eqnarray}
where $\delta\Omega_{\rm do}$ is the solid angle subtended by the disk grid
element area to the observer, $g_{\rm do}$ is the redshift factor of a photon
propagating from the disk to the observer and is given by 
$g_{\rm do}=({\bf p}_{\rm d}\cdot {\bf u}_{\rm d}/{\bf p}_{\rm o}
\cdot {\bf u}_{\rm o}$),
and $E_{\rm Fe}$ is the fluorescent line energy in the rest frame.  
The ${\bf p}_{\rm d}$ is the photon's momentum when it is emitted from 
the disk element with 4-velocity ${\bf u}_{\rm d}$ and ${\bf p}_{\rm o}$ is 
the photon's momentum when it is received 
by the observer with 4-velocity ${\bf u}_{\rm o}$ (see also similar 
definitions in Appendix B).
Note here that the illumination is not time varying since the flare is
assumed to be not an instantaneous flash but a constant source of
illumination (i.e. not time dependent).

If there are a number of flares located at different locations at a given 
time, we can use equation (5) to calculate the observed \Fe line
flux variations by summing over these flares:
\begin{equation}
F_{\rm Fe}(E)=\sum_{n}\Psi^{(n)}(E),
\end{equation}
where the superscript $n$ denotes the $n$th flare.
We use equation (6) to approximate the emission from a ring by summing
over a large number (say, 400) of discrete point sources at different
azimuthal angles (in \S~\ref{sec:resoff}). In doing so, the illumination
patterns for these discrete points evenly distributed at different 
azimuthal angles can 
be derived by rotating the illumination pattern of the one with azimuthal
angle of $0\arcdeg$.

3) We use the formulae in Appendix B to calculate the direct flux component
through propagation of photons from the X-ray source to the observer, and
hence the EW of the \Fe line. Generally, the reflected component around the
\Fe line energy band is very small ($\sim 0.1$, see Fig.~11 in George \&
Fabian 1991), and is ignored for simplicity. Thus the EW is 
given by
\begin{equation}
EW({\rm Fe}~K\alpha)\simeq\frac{\sum_{E}F_{\rm Fe}(E)\Delta E}
{F^{\rm obs}_{\rm dir}(E_{\rm Fe})},
\end{equation}
where $F^{\rm obs}_{\rm dir}(E_{\rm Fe})$ is the observed direct continuum
flux at energy $E_{\rm Fe}$.

\section{Calculation results} \label{sec:calres}

We focus on the effects of the bulk motion of X-ray flares in the corona on
the profile and EW of the \Fe line. This differs from previous studies which
considered only static sources above a disk. However, bulk motion in a
flare does not necessarily mean that the flare itself is moving 
\citep[]{bel99a,bel99b}. 
Comparing with the time-scale of a flare (see below),
the cooling time of the escaping particles (pairs or protons), which can
be immediately cooled down to Compton temperature $T_{\rm C}\sim 10$~keV,
could be very short, and they are replaced by newly created particles 
\citep[]{bel99a,bel99b}. Therefore, the dissipation region can be stationery.
The overall evolution time-scale of the magnetic configuration or a single 
flare or neighboring flares may be comparable to the Keplerian time-scale 
\citep[]{rom98}. For a maximum Kerr black hole with mass $M$ and
spin $a/M=0.998$, the Keplerian time-scale is about 
$1\times 10^4 M_7 (r/10M)^{3/2}[1+0.998(M/r)^{3/2}]$ sec,
where $r$ is the distance to the central black hole and $M_{7}$ is the black
hole mass in unit of $10^7M_{\odot}$. It is likely that the bulk velocity
and location of magnetic flares (radiatively accelerated or ejected) also
vary on this time-scale. So far, most \Fe line variability studies have been
performed on time intervals ranging from $10^4$ sec to days, which is
comparable to the flare evolution time scale for a system with typical
black hole mass $\sim 10^7-10^8M_{\odot}$. Observational 
investigation of the \Fe line reverberation mapping on shorter time-scales
is beyond the the capabilities of current satellites \citep[]{rey00}. The light
crossing time of the adopted line emission region within radius of 
$160M$ is about $8\times 10^3 M_{7}$~sec which is shorter than the overall
time-scale of magnetic configuration. Therefore, we neglect the time delay due
to the propagation of photons from the flare to the disk, and consider only
the effects of the bulk motion of flares on the time-averaged \Fe line
properties. At the reverberation mapping time-scale, the X-ray spectrum of a
flare tends to vary in a complicated way because the flare detaches from the
underlying accretion disk \citep[]{pf99}. This is in contrast to the 
instantaneous flare with fixed spectral slope adopted in previous work
\citep[]{rey99}. Thus the reverberation properties will be significantly 
affected; however, the reverberation study of an evolving flare is beyond the
scope of the present paper. 

We consider three typical bulk motions for flares: 1) on-axis flares with bulk
motion along the Kerr black hole axis; 2) off-axis flares with bulk motion 
upward or downward normal to the accretion disk plane; 3) off-axis flares with
bulk motion outward or inward along the radial direction.

\subsection{On-axis flares} \label{sec:reson}

If a primary X-ray flare moves outward with mildly relativistic bulk
velocity, the relativistic beaming effect obviously competes with
the gravitational lensing effect on both the profile and EW of the \Fe
line. On the one hand, if the flare is very close to the black hole (say,
$r<15M$), a significant fraction of the X-ray photons from the flare are
bent into the disk plane by the strong gravitation field \citep[]{mm96}. The
illumination flux onto the inner accretion disk is then enhanced whereas the
direct continuum flux is reduced. Due to the strong lensing effect, the red
wing of the line profile extends downward toward lower energies and the EW
increases \citep[]{mkm00,dl01}. On the other hand, if the flare moves with
an outward bulk velocity along the axis, direct flux will increase, and 
illumination flux impinging onto the disk surface will decrease as a result
of relativistic beaming effect \citep[]{rf97}. As a result, both the red
wing and EW are reduced.

\begin{figure}
\epsscale{0.8}
\plotone{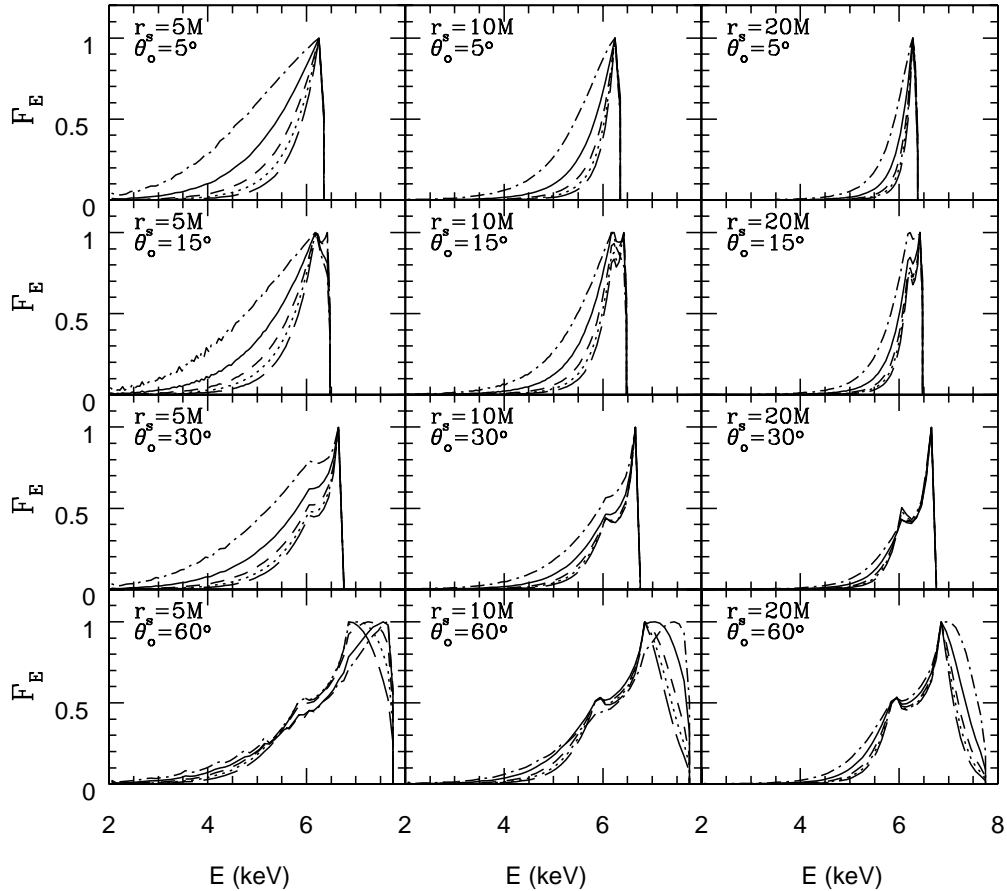}
\figcaption{Predicted line shapes for the cases of on-axis flares with
different bulk motions in a maximum Kerr black hole-accretion disk
system ($a/M=0.998$) as seen by distant observers at $1000M$
with inclination angles
of $\theta_{\rm o}=5$\arcdeg, $15$\arcdeg, $30$\arcdeg~ and $60$\arcdeg,
respectively. The on-axis flares are located at $5M$ (left column), $10M$
(middle column) and $20M$ (right column), respectively. In the
figures, the bulk velocities of flares are $-0.3$ (short-dashed-dotted lines),
$0.0$ (solid lines), $0.3$ (short-dashed lines), $0.5$ (dotted lines)
and $0.7$ (long-dashed lines). The line profiles are normalized by their
peak values regardless of their EW in order to clearly show the differences
in line profiles, since the dynamic range of the various iron lines is too
great to allow either the absolute units or a normalization by equivalent width
to be conveniently plotted (see also in Figs.~\ref{fig:offaxishape} and
\ref{fig:offaxisprof}).
\label{fig:onaxisprof}}
\end{figure}

\subsubsection{The line profile and the equivalent width} \label{sec:reson_line}

\begin{figure}
\epsscale{0.6}
\plotone{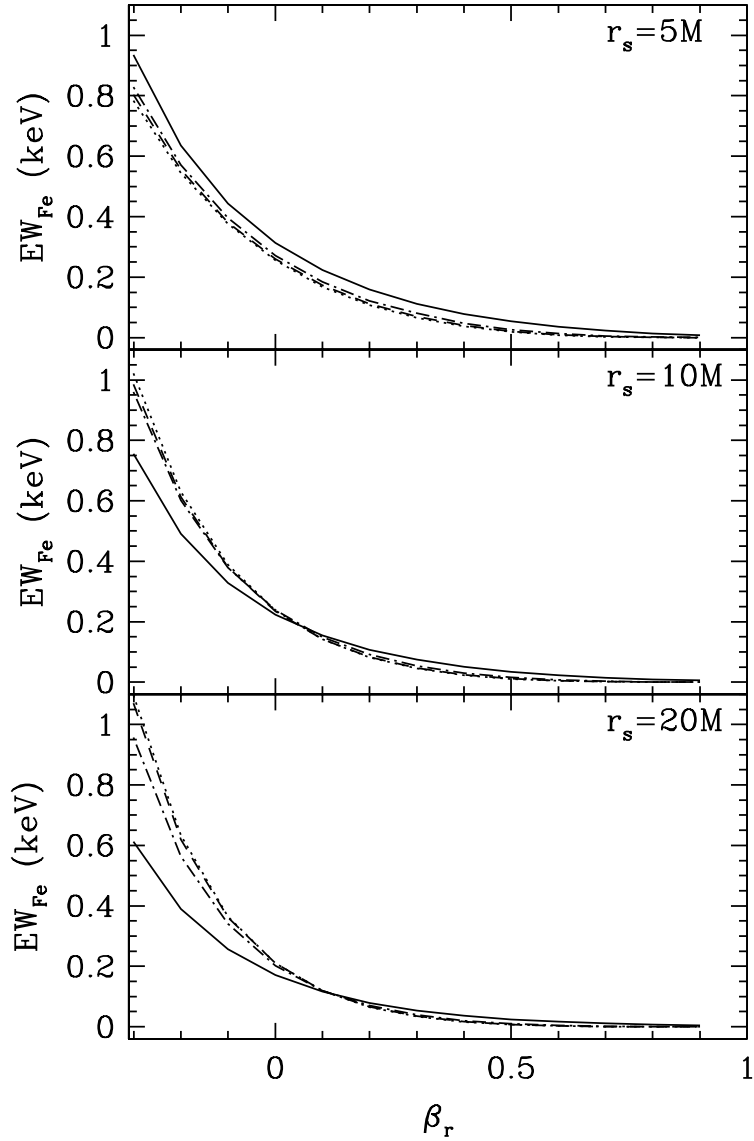}
\figcaption{The equivalent width of the \Fe line as a function of the
bulk velocities of on-axis flares. From top to bottom, the flares
are located at $r_{\rm s}=5M$, $10M$ and $20M$, respectively.
The distant observers are located at $\theta_{\rm o}=5$\arcdeg~ (dotted
lines), $15$\arcdeg~ (short-dashed lines),  $30$\arcdeg~ (short-dashed-dotted
lines) and $60$\arcdeg~ (solid lines).
Note that the dotted-line almost overlaps the short-dashed-dotted
line in all panels.
\label{fig:onaxisew}}
\end{figure}

We calculate line profiles for various cases using Monte-Carlo
simulation. The predicted line shapes and EWs are plotted in 
Figures~\ref{fig:onaxisprof} and \ref{fig:onaxisew} for a maximum Kerr
black hole with spin parameter $a/M=0.998$. Both the line profile and EW
are detected by an observer at a distance of $1000M$ with inclination
angles $\theta_{\rm o}=5$\arcdeg, $15$\arcdeg, $30$\arcdeg~ and $60$\arcdeg,
respectively. As discussed by \citet[]{mm96} and \citet[]{dl01},
the extension of the red wing, which is controlled by the strength
of the gravitational redshift, increases as the fluorescent emission is
enhanced in the region close to the black hole. As the flare gets close
to the black hole, it illuminates the inner disk more intensively and the
red wing becomes larger (see Fig.~\ref{fig:onaxisprof}). The inward (outward)
bulk motion of the flare enhances (weakens) the illumination flux in the 
inner disk region and thus results in strong (weak) flux of highly 
redshifted \Fe line emission, 
which is clearly illustrated in Figure~\ref{fig:onaxisprof}.
Indeed, the extension of the red wing shrinks as the outward bulk velocity of
the flare increases. Conversely, the red wing extends down to lower energies
as the inward bulk velocity of the flare increases (see 
Fig.~\ref{fig:onaxisprof}). For distant observers at low inclination angles
($\theta_{\rm o}\simlt 30\arcdeg$) and for flares close to the black hole
($r\simlt 15M$), the red wing of the \Fe line profiles vary substantially 
with bulk velocity. In contrast, the location of the line blue `edge', 
which is determined by the inclination angle, is insensitive to the bulk motion
of flares. {\it However, one cannot obtain a very narrow \Fe line 
for an X-ray flare close to the black hole even if the flare moves outward
with a large bulk velocity of $\sim0.5-0.7$.} For distant observers with high
inclination angles ($\theta_{\rm o}\simgt 30\arcdeg$) and/or for flares
further from the black hole ($r\simgt 15M$), the mildly relativistic bulk
motion of flares causes only small changes in the red wing. At higher 
inclination angles ($\theta_{\rm o}\sim60\arcdeg$) some variation of 
the blue `edge' is seen.
The line narrows as the bulk velocity of the flare increases.

The enhancement of the \Fe line intensity for low-height flares
($\simlt 15M$) is caused by the focusing
of the primary X-ray emission to the disk plane \citep[]{mm96,dl01}.
The relativistic bulk motion of flares is another important factor
influencing the EW \citep[]{rf97}. As seen
in Figure~\ref{fig:onaxisew}, the EW is greatly enhanced for
the downward moving flares (this enhancement can be as large as a factor
of $3-5$ for a flare with downwards bulk velocity of 0.3), while the EW
decreases with increasing flare outward bulk velocity. There is likely
a range of bulk velocities of flares (at the same location
or different locations) in a black hole$-$accretion disk system, and this may
result in a large range of EWs. For example, if on-axis flares are located
at $5M$ with bulk velocities from $-0.1$ to $0.1$, where positive velocity
means outward bulk motion and negative velocity means downward bulk motion, 
the EW can vary in range from 450eV to 200eV 
respectively (see Fig.~\ref{fig:onaxisew}).
The EW drops rapidly with increasing bulk velocity, down to very small
values ($\simlt 50$ eV) for bulk velocities $\simgt0.5$, even for
flares close to the black hole (radius $\sim 5M$).

\subsection{Off-axis flares} \label{sec:resoff}

\subsubsection{The \Fe line profiles} \label{sec:resoff_prof}

\begin{figure}
\epsscale{1.0}
\plotone{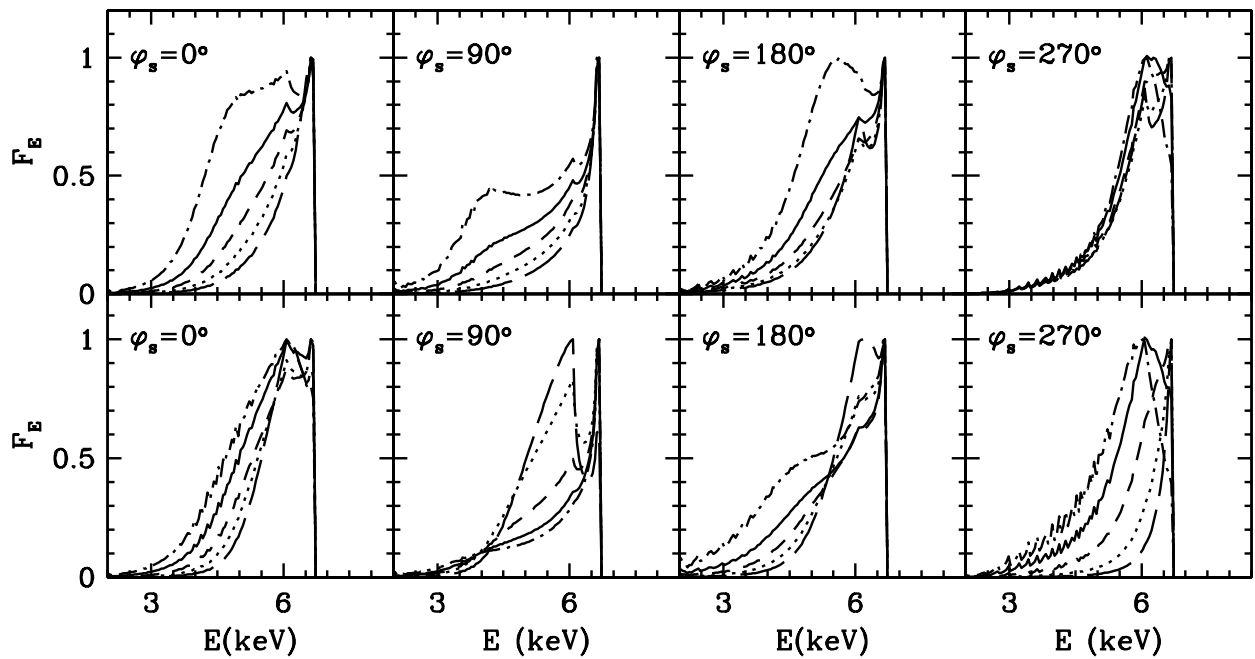}
\figcaption{ Example line profiles for a maximum Kerr black hole-accretion disk
system, as seen by a distant observer with inclination of $30\arcdeg$ and
azimuthal angle $\varphi_{\rm o}=0\arcdeg$. The primary flares are located
at $r_{\rm s}=6M$, $\theta_{\rm s}=50\arcdeg$ and $\varphi_{\rm s}=0\arcdeg$,
$90\arcdeg$, $180\arcdeg$ and $270\arcdeg$, respectively. The four top
panels are for the flares with upward/downward (along $z$ direction defined
in Appendix A) bulk motion, and the 
four bottom panels are for the flares with outward/inward bulk motion. In each
panel, the bulk velocities of flares are $-0.3$ (short-dashed-dotted lines),
$-0.1$ (solid lines), $0.1$ (short-dashed lines), $0.3$ (dotted lines)
and $0.5$ (long-dashed lines). All lines are normalized by their
peak values.
\label{fig:offaxishape}}
\end{figure}

In this section, we focus on the effects of the inward/outward
and upward/downward bulk motion of the flares. Off-axis flares with
constant angular
velocity have been considered by \citet[]{yl00} and \citet[]{rus00} for both
Schwarzschild and Kerr black hole$-$accretion disk systems. We shall not
consider the cases of orbiting flares or those\\
(co-)rotating with the disk.
Since the illumination is no longer
axisymmetric for off-axis flares, distant observers are located at 
($r_{\rm o}=1000M$, $\varphi_{\rm o}=0 \arcdeg$) for simplicity. The 
inclination angle is allowed to vary. The flare 
locations are fixed at $r_{\rm s}=6M$ and  $\theta_{\rm s}=30\arcdeg$ but
we allow the azimuthal angles $\varphi_{\rm s}$ to vary between $0$\arcdeg,
$90$\arcdeg~ (above the receding side of the disk), $180$\arcdeg~ to 
$270$\arcdeg~ (above the approaching side of the disk).
Example line profiles are presented in
Figure~\ref{fig:offaxishape}. We only show line shapes seen by
a distant observer at an inclination of $30\arcdeg$ because the dependence
of the line shapes on the inclination is similar to the case of
on-axis flares. The top panels in Figure~\ref{fig:offaxishape} show that the
\Fe line width generally narrows and the red part of the line 
weakens with increasing upward bulk velocity (unless the flares are located
above the approaching side of the disk). In case of flares located above the
approaching side of the disk ($\varphi_{\rm s}=270\arcdeg$),
the difference in the line profiles is small even when the difference in the
bulk velocity of the flares is relatively large (say, $0.5$). This is because
the receding side of the disk is much less illuminated and correspondingly 
fewer highly redshifted \Fe line photons are emitted from the receding side.
For flares with outward/inward bulk motion (along the radius), the red wing
of the line shrinks with increasing outward bulk
velocities of flares except where the flares are located above the receding
side of the disk (see the bottom panels in Figure~\ref{fig:offaxishape}). 
For flares located above the receding side of the disk 
($\varphi_{\rm s}=90\arcdeg$), the illumination of the inner region of the
approaching side of the disk decreases significantly, while the illumination
of the receding side of the disk decreases less as the outward bulk velocity
increases. Thus in comparison with the red part of the \Fe line, the blue
part weakens with increasing outward bulk velocity and a sharp 
red peak emerges around 6~keV when the bulk velocity is large ($\ga 0.5$, 
see Fig.~\ref{fig:offaxishape}). 
However, the maximum extension of the red wing for flares with small outward
bulk velocity does exceed that for flares with large outward bulk velocities.  

\begin{figure}
\epsscale{1.0}
\plotone{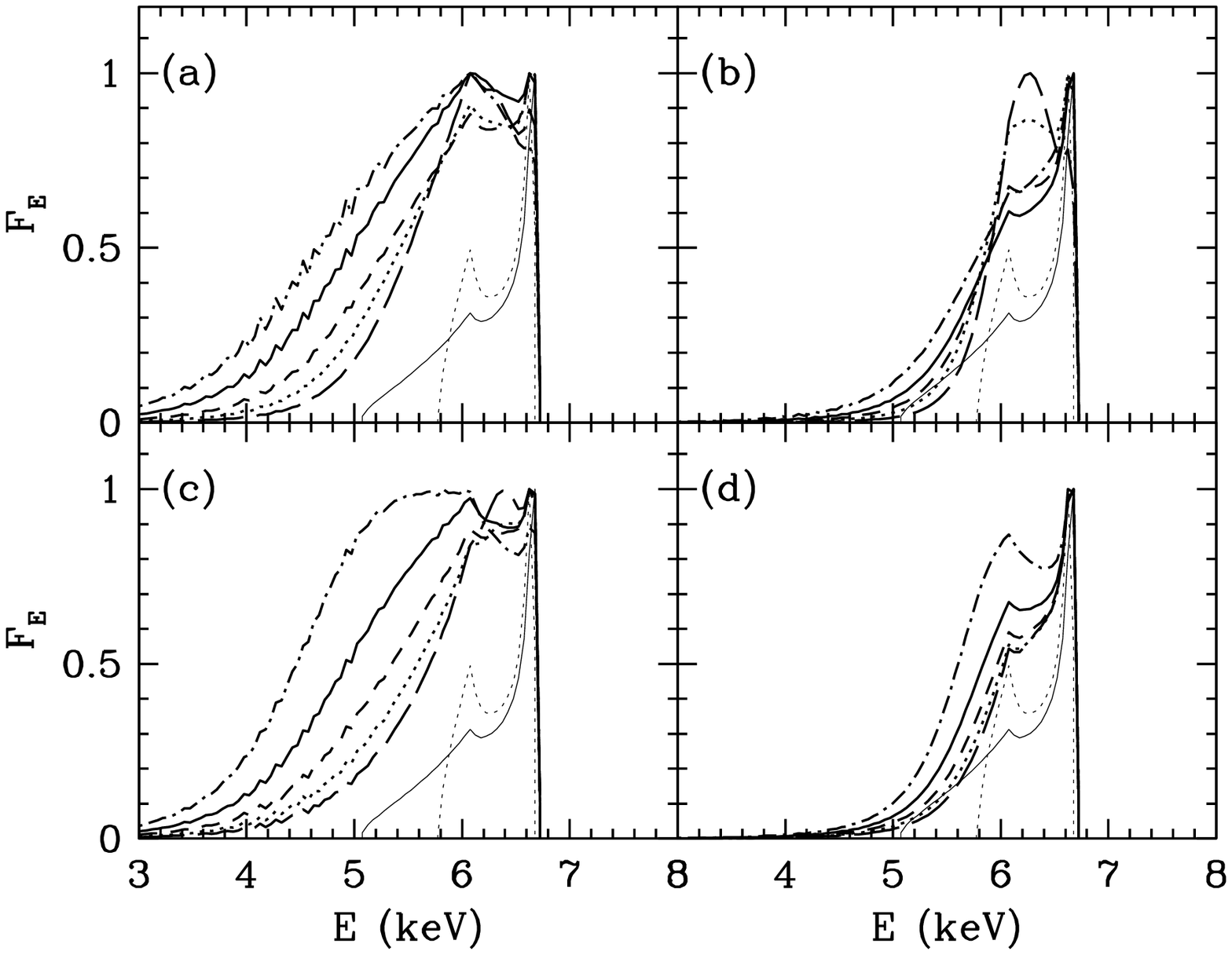}
\caption{ Line profiles for X-ray emitting rings with bulk motions.
The observers are fixed at $r_{\rm o}=1000M$, $\theta_{\rm o}=30\arcdeg$. 
The X-ray emitting rings are located at $r_{\rm ring}=6M$,
$\theta_{\rm ring}=50\arcdeg$ (left panels) and 
$r_{\rm ring}=15M$, $\theta_{\rm ring}=50\arcdeg$ (right panels).
The top panels are for rings with upward/downward bulk velocities 
$-0.3$ (short-dashed-dotted lines), $-0.1$ (solid lines), $0.1$ (short-dashed
lines), $0.3$ (dotted lines) and $0.5$ (long-dashed lines). The bottom panels
are for rings with inward/outward bulk motion. In all four
panels, the light lines represent the best-fitted \Fe line profiles for
{\it ASCA} observations of NGC5548 \citep[light solid line]{chiang99} and
IC4329A 
\citep[light dotted line]{dmz00}. The lines are normalized at
peak values regardless of their EWs.
\label{fig:offaxisprof}}
\end{figure}

To mimic a continuous corona with bulk motion, we first consider the cases
of X-ray emitting rings with bulk motion atop the accretion disk. The line 
spectra are presented in  Figure~\ref{fig:offaxisprof} (a), (b) (upward/downward
bulk motion) and in Figure~\ref{fig:offaxisprof} (c) and (d) (inward/outward
bulk motion) for ring-like X-ray emitting sources located at ($r_{\rm ring}$, 
$\theta_{\rm ring}$); these are approximated by summing over
a large number (400) of discrete point sources at different azimuthal
angles using equation (6). Only line shapes observed by a distant 
observer with an inclination of $30\arcdeg$ are shown. As expected,
for the flares with larger outward or upward bulk velocities, the red wing
is less extended, ie. the \Fe line is narrower than that for flares
with smaller velocities. The location of the blue `edge', which is controlled
by the inclination, does not vary with the bulk motion velocities of the
ring. As seen from Figure~\ref{fig:offaxisprof} (a) and (c), it is difficult
to get a very narrow \Fe line if the X-ray emitting ring is close to the
black hole, unless the bulk motion of the ring is highly relativistic.
Note also that the EW is negligible if the X-ray emitting ring has
a highly relativistic bulk motion (see Fig.~\ref{fig:onaxisew} and
also Figs.~\ref{fig:offaxiszew}, \ref{fig:offaxisrew}, 
\ref{fig:offaxiszew70} and \ref{fig:offaxisrew70}). 

If a large number of magnetic flares occur randomly above the inner disk 
region at a given time, the dissipation region can be considered as a 
continuous corona. The \Fe line is then the summation of those lines produced
from X-ray emission
rings at different locations weighted by their intrinsic luminosity. The \Fe
line for a corona with upward/outward bulk motion is inevitably narrower 
than that for a corona without bulk motion if the majority of the X-ray 
emission comes from the inner region. If the X-ray emission is dominated
by a single flare or neighboring flares induced by avalanche model
at any given time \citep[]{pf99}, non-axisymmetric illumination should be
important. Thus, the \Fe line can be either very narrow or very broad
depending on different flare locations (e.g. Fig.~\ref{fig:offaxishape}). 
However, the average line profile
over several flare events should be similar to that for a continuous corona.

\subsubsection{The intensity and equivalent width of the \Fe line} \label{sec:resoff_ew}

\begin{figure}
\epsscale{1.0}
\plotone{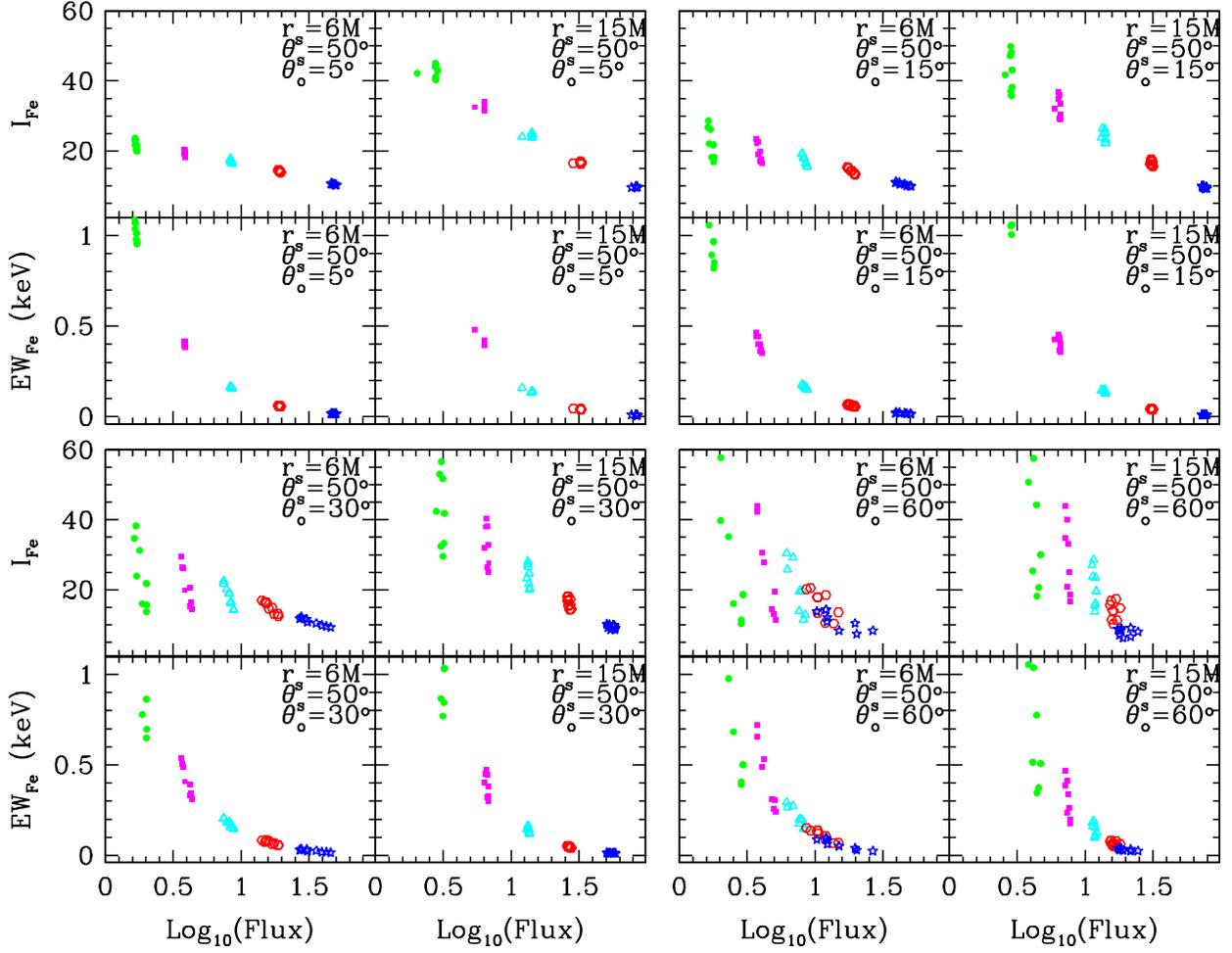}
\caption{ 
The intensity (I$_{\rm Fe}$) and equivalent width (EW) of
the \Fe line as a function of the direct flux for flares with different 
upward/downward (along $z$ direction defined in Appendix A) bulk velocities
$\beta=-0.3$ (solid circles), $-0.1$ (solid squares), $0.1$ (open triangles),
$0.3$ (open hexagons) and $0.5$ (open stars). Solid symbols represent
downward (or inward in Figs.~6 and 8) bulk motion, while open symbols
represent for upward (or outward in Figs.~6 and 8) bulk motion. The flare
location and
the observer inclination are labeled in each panel. Note that the eight
points for each point style in each panel represent the flare azimuthal
angles of $0\arcdeg$, $45\arcdeg$, $90\arcdeg$, $135\arcdeg$, $180\arcdeg$,
$225\arcdeg$, $270\arcdeg$ and $315\arcdeg$, respectively.
The intensity of the \Fe line and the direct flux are in arbitrary units.
All the flares are assumed to
have the same X-ray intrinsic luminosity and spectra.
\label{fig:offaxiszew}}
\end{figure}

\begin{figure}
\epsscale{1.0}
\plotone{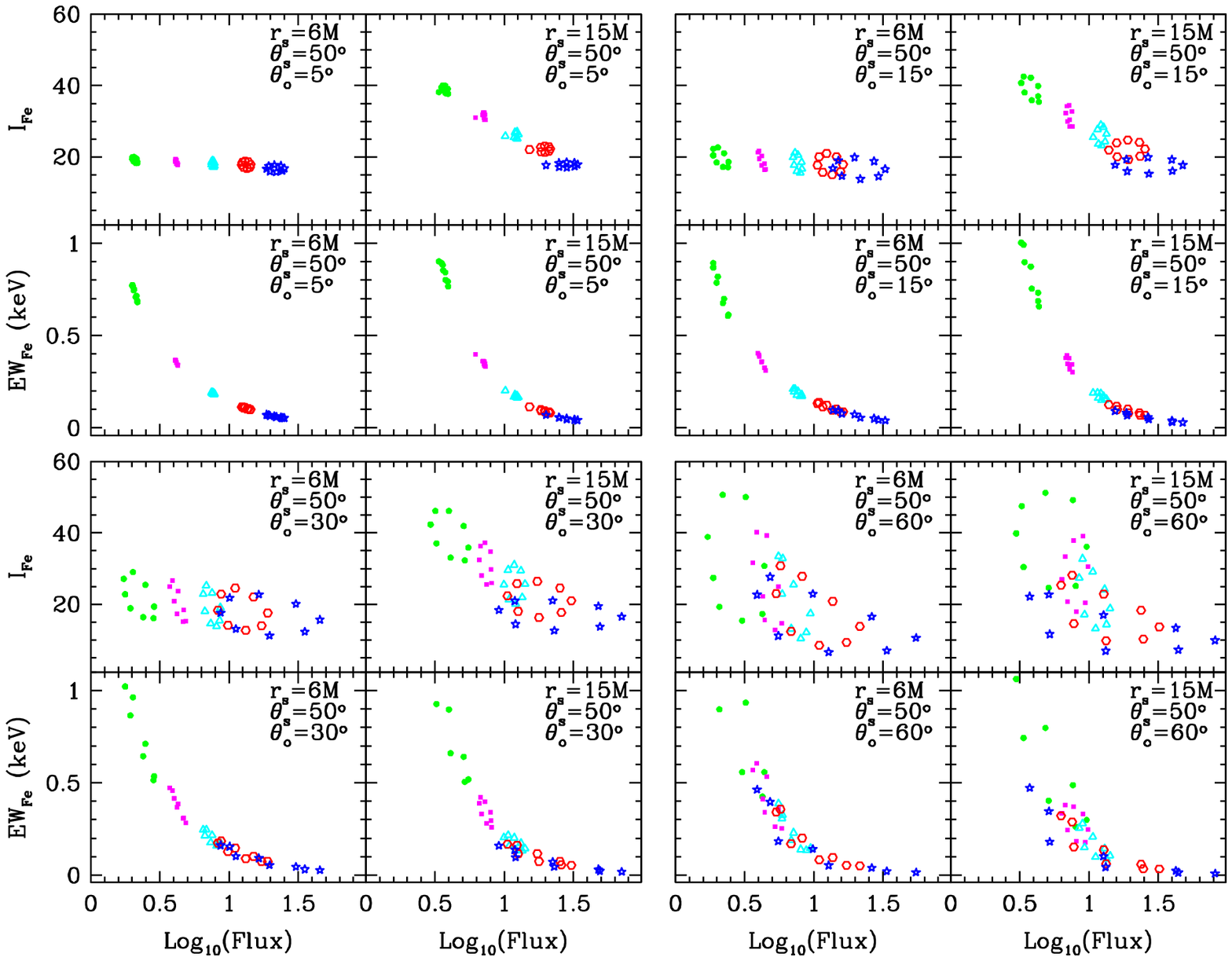}
\caption{ Legend as for Fig.~\ref{fig:offaxiszew} but for flares with
inward/outward bulk motion (along $r$ direction).
\label{fig:offaxisrew}}
\end{figure}

\begin{figure}
\epsscale{1.0}
\plotone{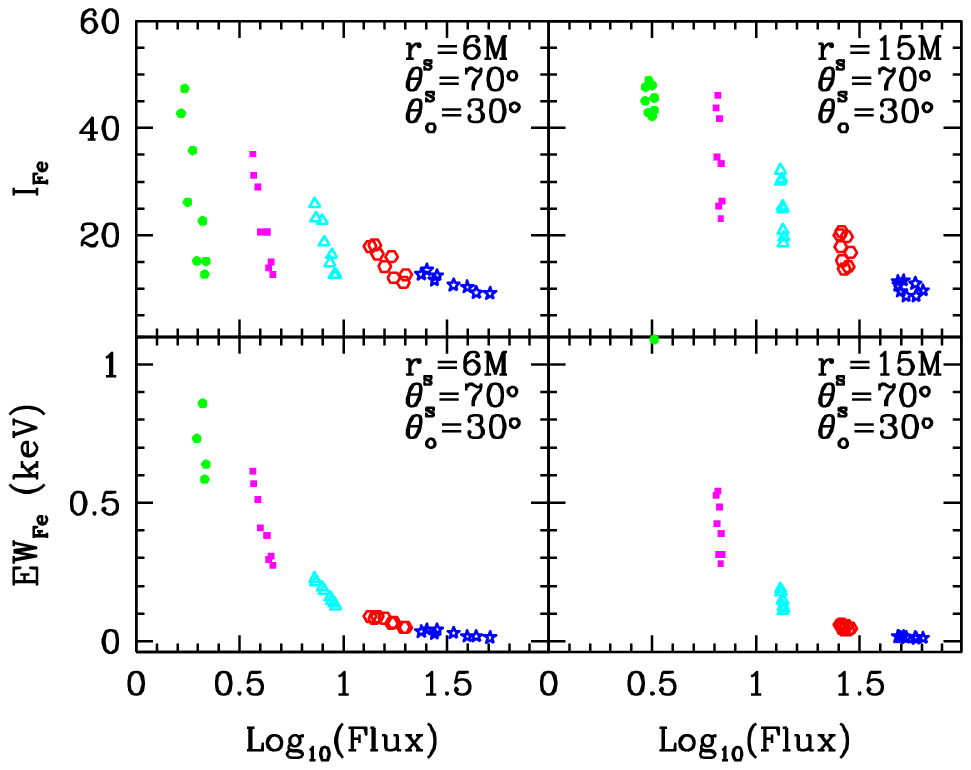}
\caption{ Legend as for Fig.~\ref{fig:offaxiszew} but for flares with
$\theta_{\rm s}=70\arcdeg$ and only for a observer with inclination of
$30\arcdeg$.
\label{fig:offaxiszew70}}
\end{figure}

\begin{figure}
\epsscale{1.0}
\plotone{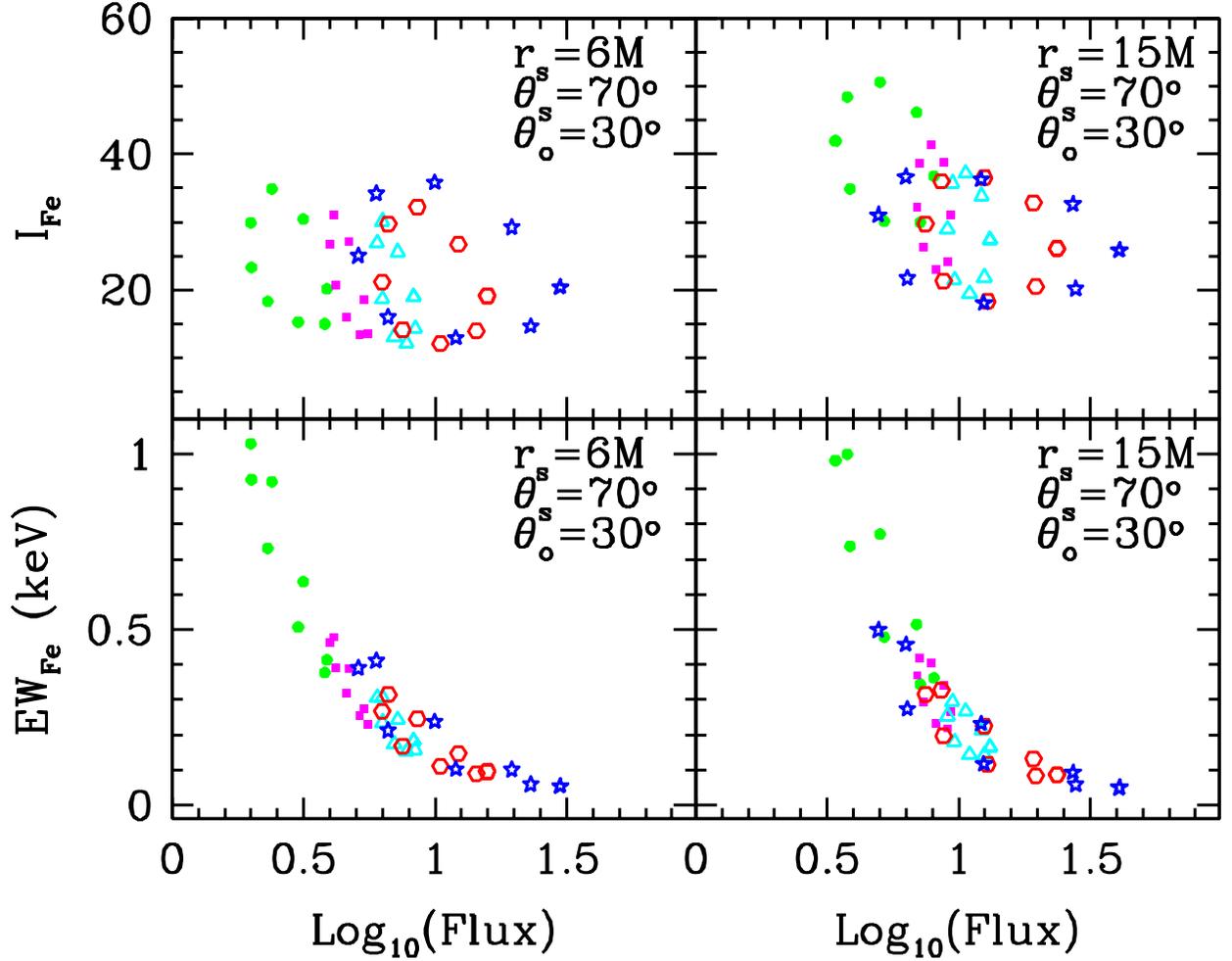}
\caption{ Legend as for Fig.~\ref{fig:offaxiszew70} but for flares with
inward/outward bulk motion (along $r$ direction).
For each point style, generally the point with the largest I$_{\rm Fe}$
corresponds to the azimuthal angle of $\varphi_{\rm s}=270\arcdeg$~ (on the
approaching side of the disk),
and with the symbols moving counterclockwise (or clockwise), the
azimuthal angles change from 270\arcdeg, 315\arcdeg, 360\arcdeg~ (or 0\arcdeg),
45\arcdeg, 90\arcdeg, 135\arcdeg, 180\arcdeg, to 225\arcdeg~ for
downward/inward (or upward/outward) flares.
\label{fig:offaxisrew70}}
\end{figure}

We contrast the intensity and EW of
the \Fe line with the direct flux from the flares with different location
and bulk motions, but same intrinsic X-ray luminosity:
Figures~\ref{fig:offaxiszew} and \ref{fig:offaxiszew70} show upward/downward
bulk motion, and Figures~\ref{fig:offaxisrew} and \ref{fig:offaxisrew70}
inward/outward bulk motion along the radius. In cases of flares with upward
or outward bulk motions, the continuum flux increases with increasing bulk
velocities of flares due to the beaming effect. {\it It is interesting to note
that a significant change in the direct flux does not necessarily mean a
corresponding change in intrinsic X-ray luminosity. Instead, this change
can be due to the change of the location and bulk velocity of flares.
Moreover, a weak flare with a large bulk velocity towards the observer
may appear as a bright flare, whereas a luminous flare closer to the
central black hole with a small bulk velocity towards the observer may appear
as a minimum period in the light curve.} If a flare is very close to the black
hole (with the radius $\simlt 10M$; for example, $6M$ in
Figures~\ref{fig:offaxiszew} and \ref{fig:offaxisrew}), the number of
photons impinging onto the disk decreases with increasing bulk velocity of
the flare, but the trend is much weaker than the case of a flare located away
from the black hole because of the dominance of the gravitational lensing 
effect over relativistic beaming. Therefore general relativistic effects cause
only small difference in the total \Fe line flux for the most inner flares
with different bulk velocities. However, the EW decreases rapidly with
increasing bulk velocities of flares due to the rapid increase of the direct
flux. If a flare is located above the approaching side of the disk, the
total line flux is enhanced due to the Doppler effect. Conversely, the total
line flux is repressed if the flare is located above the receding side of 
the disk. This asymmetric effect introduces a large scatter in the \Fe line
flux and EW, especially for observers with large inclination angles and
flares with large bulk velocities. For out-flowing flares, the direct flux 
is significantly affected by the angle between the flare bulk velocity vector
and the line of sight. The direct flux is significantly enhanced for
a flare moving towards the observer, i.e. located above the part of the
disk facing the observer.

The most intriguing cases have flares with inward/outward bulk
motions (e.g. Fig.~\ref{fig:offaxisrew70}) in which {\it the total \Fe line
fluxes are almost independent of the bulk velocity of flares (close to the
black hole with the same intrinsic luminosity), and consequently the EW is
inversely proportional to the continuum flux, for distant observers with
inclination angles $\simlt 30\arcdeg$}. It can be clearly seen in 
Figures~\ref{fig:offaxiszew} --- \ref{fig:offaxisrew70} that a constant
\Fe line flux can arise only for flares with low height above the disk and
outward/inward (along the radius) bulk velocity. \footnote{To make this
easier to understand, consider an extreme case of a flares motion parallel
to the infinite disk plane ($\theta_{\rm s}\sim 90\arcdeg$). The number of
reflected photons will be roughly
a constant whatever the bulk velocity, without considering general 
relativistic effects. A similar result can be obtained even for flares with
$\theta_{\rm s}\sim 50\arcdeg$ in the very inner region if general relativistic
effects are included.} This requirement is possibly satisfied
in realistic cases since it has been argued that the height of luminous
magnetic flares cannot be much more than a few times the pressure height above a
thin accretion disk \citep[]{nk01}.

\section{Applications and Discussion} \label{sec:app}

\subsection{Narrowness of the line profile} \label{sec:app_nar}

If only the effects of the bulk motion of flares on the profile and EW of
the \Fe line are considered, a larger EW is found to be correlated with a
more extended red wing of the \Fe line (see Figs.~\ref{fig:onaxisprof} and
~\ref{fig:onaxisew}). This tendency is roughly consistent with observations:
the larger the line EW, the more extended the red wing \citep[e.g.][in the
parameters obtained from the line modeling, a larger EW is 
accompanied by a smaller inner radius]{lz01}.
Therefore, bulk motion in the corona can be introduced as a possible mechanism
for producing the `narrowness' of the \Fe line seen in some Seyfert 1
galaxies (see below); this is in addition to the generally
introduced parameter of the truncation radius of the cold disk
[the cold disk is truncated at a larger radius rather than the radius at the
marginal stable orbit due to the transition of the inner disk to a hot
geometrically thick and optically thin
accretion flow or the high ionization of the disk material within this
radius \citep[and references therein]{lz01}]. It has been suggested that
outflow/inflow in the corona provides a possible explanation for the
$R-\Gamma$ relation \citep[]{bel99b}. 
With the calculation in \S~\ref{sec:reson_line}, this
out-flowing magnetic flares model is expected to consistently account for
the `narrowness' of the \Fe line profile, the variation of the EW, and the
$R-\Gamma$ relation in some Seyfert 1 galaxies \citep[]{bel99b,zls99,lz01}.
Note that the ``narrowness'' of the line also depends on the location of
the flare as seen in Fig.~\ref{fig:onaxisprof}.

The model of a point source of X-ray emission along the rotation axis of
the accretion disk is only an approximation to an accretion disk corona.
Perhaps, a standard corona with upward/downward (or inward/outward) bulk 
motion is better approximated by a ring as shown in top (bottom)
Fig.~\ref{fig:offaxisprof}. Even with bulk velocity of 0.5 the line appears
relatively broad with a red-wing extending down below 5~keV if the ring is
located at a small radius, e.g. $6M$ (see Fig.~\ref{fig:offaxisprof} left
panels). Here, one important question is: can an overall upward or outward
bulk motion in the corona account for the `narrower' \Fe line found in some
Seyfert 1 galaxies?  \citet[]{dmz00} found that IC4329A has a relatively 
narrow line which can be modeled using a disk line having an inner truncation
radius of about $50M$ and EW of $210\pm45$eV. NGC4593 was
also found to have a relatively narrow \Fe line with an inner truncation
radius of $30M$ \citep[]{lw00}. Even in NGC5548, the \Fe line
emission region was found to be truncated at an inner radius of $15M$
\citep[]{chiang99}. The best-fit line profiles for NGC5548 and IC4329A
are also shown (in Fig.~\ref{fig:offaxisprof}) as light solid and light
dotted lines, respectively. A significant fraction of the X-ray emission may
come from the region within $15M$ in the thin disk$-$corona regime,
which would indicate that the bulk motion in the corona is not the dominant
factor making the line `narrower' in some objects. As shown in
Figure~\ref{fig:offaxisprof}, obviously the observed
`narrower' line profile in IC4329A cannot be interpreted only by the corona
bulk motion in this case. Other parameters (possibly the generally introduced
inner truncation radius caused by the total ionization or disruption
of the disk within this radius) are also still needed, or the corona in
IC4329A is more extended and most of the X-ray emission comes from
an outer region with radius of $\sim 50M$.
For NGC5548, a narrow \Fe line which may be emitted from the broad line region
or molecular torus, has been detected with the {\it Chandra} High-Energy
Transmission Grating \citep[]{yaq01}. The corona bulk motion may account for
the `narrower' \Fe line in NGC5548 by removing the contribution from distant
material. Here, even the X-ray emission may come mainly from the region
around or within $15M$, and the ionization or disruption of the disk
may not be necessary. Detailed models are beyond the scope of this paper.

\subsection{The redshifted \Fe line from radio-loud quasars} \label{sec:app_rl}

With the observation of the Calibration and Payload Verification (Cal-PV)
phase of {\it XMM$-$Newton}, a redshifted \Fe line at 6.15keV (quasar
frame) has been found in the radio-loud quasar PKS0537-286 (redshift
$z=3.104$) with
an EW of 30eV, which suggests the existence of cold matter near the central
black hole \citep[]{reev01}. The overall spectral energy distribution of
this object is dominated by a flat power-law X-ray emission with a spectral
index of $\alpha\sim 0.27$, indicating that the dominant X-ray emission
mechanism is the inverse Compton emission associated with a face-on
relativistic jet \citep[]{reev01}. The reflection fraction was constrained
to be $R\simeq0.25$ indicating the reflection material subtends a solid-angle
much lower than the $2\pi$ expected from an accretion disk. \citet[]{reev01}
explained these features using two continuum emission components: a
`Seyfert-like' steep power-law component originating from the region near the
accretion disk and associated with its reflection continuum, and a hard bright
jet component with an X-ray flat slope. Generally, the radio jet is believed
to be launched from the region close to the central black hole, which is
possibly related to the ejecting X-ray plasma. These X-ray features in
PKS0537-286 may also be explained in a unified model: the X-rays are emitted
from the upward ejecting plasma near the central black hole (with radius of
tens $M$) having an effective bulk velocity of $\sim 0.5-0.7$. The downward
X-ray photons from the plasma are reflected by the accretion disk and thus
produce the \Fe line with a small EW and the small reflection fraction due to
the beaming effect. The redshifting of the line (from 6.4~keV to 6.15~keV)
could also be explained as due to transverse-Doppler effect and gravitational
redshift if this quasar is seen at a small inclination angle [radio-loud 
quasars are
generally believed to be seen at a small inclination angle \citep[]{up95},
e.g. $\theta_{\rm o}\la 15\arcdeg$, see also the upper panels of
Fig.~\ref{fig:onaxisprof}]. This X-ray emission region could be the base of
the launching jet. Therefore, the \Fe line together with its reverberation
could be a useful tool to probe the region of jet formation and to measure
the initial velocity of the jet as well as the inclination of the disk in
radio-loud quasars. This may reveal the physical connection between the jet
properties and the cold accretion flow using future X-ray satellites 
({\it Constellation X-ray Mission} and {\it XEUS}).

\subsection{Explanation of the constant line flux in 
some objects} \label{sec:app_app}

It is still a puzzle that the \Fe line flux appears to be constant in spite
of strong continuum variation, and that the EW is anti-correlated with the 
continuum flux as in MCG-6-30-15 \citep[]{lee00} and NGC5548 
\citep[]{chiang99}. This was explained by the presence
of a hot ionized skin of an accretion disk induced by thermal instability
\citep[]{nkk00} or the resonant trapping followed by Auger destruction in
an ionized disk which suppresses the emission of the \Fe line
\citep[]{brf01}. However, ionization physics
in the inner disk is unlikely to explain the behavior of \Fe line
variation, at least in MCG-6-30-15 since the line profile suggests
the existence of cold material within a radius of 
$5M$ \citep[]{iwa99,zr01}.
The variation of the \Fe line profile in MCG-6-30-15 (and possibly also in
NGC5548) may be explained by the out-flowing magnetic flares model in which
the bulk motion and location of flares may vary [e.g. as discussed by 
\citet[]{iwa99}]. A
subsequent question is: can the bulk motion and location of magnetic flares
consistently explain the pattern of the variation in line intensity and EW
corresponding to significant variations in continuum flux?

On occasion, the X-ray emission is dominated by a single (or a small
number of) large flare(s) or neighboring flares \citep[e.g.][]{pf99};
but typically the X-ray emission may be a sum of a dozen or so overlapping
flares. In the special case where the X-ray emission in an object at any 
given time is dominated by a
single flare or neighboring flares in the inner region ($r\simlt 15M$),
a significant non-axisymmetric illumination of the accretion disk is
produced. If flares produced at different time have similar intrinsic
X-ray luminosity, but move outward/inward with different bulk velocities,
then the resulting \Fe line flux appears constant while the observed 
continuum flux undergoes rapid variability due to the beaming effect to a 
observer with a inclination $\la 30\arcdeg$ (see Fig.~\ref{fig:offaxisrew70}).
It is obvious that a constant \Fe line flux can also be obtained from a 
continuous corona (or a large number of overlapping flares) with varying
bulk velocity in the region ($r\simlt 15M$, axisymmetric illumination) by 
averaging the symbols for the flares with same bulk velocity
but different azimuthal angles in Fig.~\ref{fig:offaxisrew70}. Therefore,
the model of out-flowing magnetic flares qualitatively explains the
observational behavior of the \Fe line in MCG-6-30-15 \citep[]{lee00} and 
NGC5548 \citep[]{chiang99}. These two objects are both constrained to have
inclinations of $30\arcdeg$ from fits of the \Fe line profile with a
relativistic disk line. The \Fe line in MCG-6-30-15 shows dramatic
variability: it was narrow during a high-flux state and very broad
during a deep-minimum flux, but the total line flux seems to be constant
during the long {\it ASCA} observation in 1994 \citep[]{iwa96}; however, its
blue part is shifted
well below 6.4 keV during a short bright period in the long {\it ASCA}
observation in 1997 \citep[]{iwa99}. Interestingly, the \Fe line profile
has a huge red tail in the deep-minimum spectrum, and the line profile is
without the 6.4 keV component during the bright period. These observations
suggest that the X-ray source is very close to the central
black hole and probably within radius of $6M$. If most of
the X-ray is indeed emitted from the inner region with a typical radius of
$6M$ in MCG-6-30-15, then a fluctuation of the bulk velocity of flares in
the range $-0.1$ to $0.1$ (see Fig.~\ref{fig:offaxisrew70}), together
with the non-axisymmetric illumination due to the difference in the
location of flares (this non-axisymmetric illumination should be averaged 
in a long time interval) can account for the observational behavior of
the \Fe line \citep[]{lee00}.
Characterizing the \Fe line as a narrow plus a broad component, \citet[]{iwa96}
found that the intensity of the narrow component correlates with the continuum
flux, whereas the intensity of the broad component possibly anti-correlates
with the continuum flux. This observational result is also qualitatively
consistent with the out-flowing magnetic flare model. The X-ray emission in
NGC5548 is also required --- from the \Fe line profiles --- to come from the
inner egion around $15M$. However, the emission is more extended than that
in MCG-6-30-15. The observed behavior of the \Fe line in NGC5548 
\citep[]{chiang99} may also be explained by the X-ray emission being
dominated by moving magnetic flares
with bulk velocity fluctuating around $0.2$ or $0.3$ at the region around 
or within $15M$. 

It is interesting to note that if this model is correct it should
apply in many different sources, including those observed at higher
inclination angles. For example, in objects with high inclinations,
i.e.  Seyfert 2 galaxies, the \Fe line flux may not appear as a ``constant''
as in MCG-6-30-15 and NGC5548; instead, it may decrease with increasing
continuum flux (see Fig.~\ref{fig:offaxisrew}).
The \Fe line flux may also not appear as a ``constant'' but
anti-correlate with the continuum flux (see Fig.~\ref{fig:offaxiszew})
in radio loud objects (e.g. quasars, which are generally believed to be
observed at a small inclination angle) in which the out-flowing is upwards.
The outflow/inflow mechanism can vary between sources
observed at the same inclination angle (e.g. the radial outflow velocity
may scale with the accretion rate unit in Eddington limit) as suggested
above for MCG-6-30-15 and NGC5548. The case with a larger bulk
velocities may have a smaller line EW, and also a narrower line profile
due to a possible larger dissipation region.

\subsubsection{Caveats and realities} \label{sec:app_cav}

Quantitatively, it is difficult to use the above model to explain the 
observational behavior of a constant \Fe line flux in MCG-6-30-15 and
NGC5548 because the flare formation mechanism is not well known. (However,
it is also a problem for other models, e.g. the totally ionized inner disk 
or the truncation of a thin disk in the inner region, since the formation 
of the X-ray corona or the dynamics of the truncation is not fully understood.)
One major simplification in our calculation is that all flares are
assumed to have the same intrinsic X-ray luminosity. However, the intrinsic 
X-ray luminosity of different flares may be different in realistic cases. 
Unfortunately, how flares are accelerated is not clear, as is the dependence
of the bulk motion of flares on its intrinsic power. It is not yet known 
whether flares with larger bulk velocities tend to have larger intrinsic 
luminosity. The above model is viable only if the difference in the intrinsic
power for flares with different bulk velocities is small (e.g., less than a 
factor of $1.5$). This may be satisfied for the avalanche-produced neighboring
flares. Furthermore, the properties of flares may depend on their location.
For example, the innermost flares may be more energetic or have a large
bulk velocity than flares in the outer region. It is also
possible that the innermost flares move with small outward bulk
velocity due to the dominance of the gravitational field. 

The spectral indices $\alpha$ of the X-ray continuum emitted from flares with
different bulk motion and location should be
different rather than the uniform value of $1.0$ adopted in our calculation.
This difference will introduce some scatter in the EW and intensity of the \Fe
line. This scatter will be less than 10 per cent for the observed
variance of $0.2$ in the spectral index of 
MCG-6-30-15 \citep[$\alpha \sim 0.8-1.0$,][]{lee00}
and NGC5548 \citep[$\alpha \sim 0.75-0.95$,][]{chiang99} \citep[]{gf91}. 
It is not yet clear how the spectral index of the X-ray continuum from a 
flare depends on its bulk velocity and location.
Without considering general relativistic effects, the photon index of the
spectrum from magnetic flares with upward/downward bulk
motion decreases with increasing bulk velocity \citep[]{bel99a,bel99c,mbp01}.
In an individual object, different low-height flares at different
time may move with different outward/inward bulk velocities.
The spectral variability is controlled by the parameters
of temperature and optical depth of the Comptonizing medium, and the 
soft photons from the local disk radiation. It is possible that the
temperature of the disk region where the flare is born is lower than that
of the region surrounding it and less efficient at producing soft seed photons. 
So, {\it in an individual object the flare with a higher outward bulk 
velocity may
intercept more soft photons from the disk region surrounding the area where
it is born since it moves away from the latter area more quickly, and 
will thus emit a harder continuum}. Therefore, the fact that continuum 
spectra steepen with increasing continuum flux in MCG-6-30-15 \citep[]{lee00}
and NGC5548 \citep[]{chiang99} may also be consistent
with the out-flowing magnetic flare model used to explain the constant \Fe 
flux in these objects. Since it has been shown that systematic errors exist 
in the data fitting procedure of $R$ and $\Gamma$ \citep[]{ve01,nan00}, we 
do not try to explain the relation between $R$ and $\Gamma$ in MCG-6-30-15
and NGC5548. Other than the above complications, the difference in the bulk 
motion and location of magnetic flares could be the main cause of the 
observed constant \Fe line flux. Further relativistic calculations
of the X-ray spectra from the flares with bulk motion near the central black
hole are needed.

The behavior of a constant \Fe line flux in the presence of strong continuum
variation can result from the combination of the effects of a
strong gravitation field and relativistic beaming in the out-flowing 
magnetic flare model. This is only true for those objects in which the 
X-ray emission is concentrated in the innermost region with typical radii
of $\simlt 15M$ and probably in objects where the \Fe line is also broad. 
The X-ray emission in some objects (e.g. IC4329A and NGC4593) must be 
emitted mainly in a region much larger than that in MCG-6-30-15 due to the 
observed `narrower' \Fe line profile. Thus the expected variability of the
\Fe line will not be similar to that in MCG-6-30-15 and NGC5548. If the 
behavior of constant \Fe line fluxes found in MCG-6-30-15 and NGC5548
is caused by ionization physics in the disk \citep[]{rfy99,nkk00} rather
than the relativistic effects described above, then similar variability
would also be expected in those object with `narrower' \Fe line.
It is also possible that the model of out-flowing
magnetic flares works for the objects like MCG-6-30-15 with neutral disk, but
that the ionization model works for those objects
with highly ionized accretion disks.
We have to await future observations and detailed theoretical considerations
to distinguish the out-flowing magnetic flare model from the model of
highly ionized accretion disk.

\section{Conclusions} \label{sec:con}

We have performed fully relativistic calculations to reveal the effects of
the bulk motion and location of magnetic X-ray flares above an untruncated
accretion disk on the \Fe line
properties, including the \Fe line profile, the EW, and the total line
flux. The main conclusions are:

\begin{itemize}
\item{The bulk motion of magnetic flares in the corona affects the 
\Fe line profiles. The red wing of \Fe line becomes less extended as
the bulk velocity of the out-flowing magnetic flares increases. This
effect is only significant for flares close to the central
black hole.}

\item{The EW of the \Fe line drops rapidly with increasing bulk velocity
of the magnetic flares. In the out-flowing magnetic flare model, if the
X-ray emission in an object is dominated by magnetic flares in the inner
region ($r\simlt 15M$), it is difficult to get the `narrower'
lines observed in IC4329A, and a compatible EW. This suggests
that the bulk motion cannot be the dominant factor causing
the `narrowness' in some AGNs.}

\item{A fluctuation in the bulk velocity of the low-height out-flowing
magnetic flares
can account for the behavior of constant \Fe line fluxes and strong
continuum variation observed in MCG-6-30-15 and NGC5548. This is especially
interesting for the case of MCG-6-30-15, in which the \Fe line behavior is
unlikely to be caused by the ionization of the accretion disk. }

\end{itemize}

\section*{Acknowledgments}
We are grateful to Neta Bahcall and Stuart Swyither for a careful reading of
the manuscript with many helpful comments. We also thank an anonymous referee
for many helpful comments and suggestions.
YL acknowledges the hospitality of the Department of Astrophysical Sciences,
Princeton University.



\appendix
\section{Motion Constants:}
\label{appendix:motion constant}
The line element of the Kerr metric can be expressed in the form
\citep[]{chandra}
\begin{equation}
ds^{2}=-{\exp}^{2\nu}(dt)^{2}+{\exp}^{2\psi}(d\varphi-\omega dt)^{2}+{\exp}^{2\mu_{2}}
dr^{2}+{\exp}^{2\mu_{3}}d\theta^{2},
\end{equation}
where
\begin{eqnarray} \label{eq:system1}
\exp^{2\nu}&=&A^{-1}\Delta\Sigma, \nonumber\\
\exp^{2\psi}&=&A \Sigma^{-1} \sin^2\theta, \nonumber \\
\exp^{2\mu_{2}}&=&\Delta^{-1} \Sigma, \nonumber \\
\exp^{2\mu_{3}}&=&\Sigma, \\
A&=&(r^2+a^2)^2-a^2 \Delta \sin^{2}\theta, \nonumber \\
\Delta&=&r^2+a^2-2r, \nonumber \\
\Sigma&=&r^2+a^2 \cos^2\theta, \nonumber \\
\omega&=&\frac{2ar}{A}, \nonumber
\end{eqnarray}
and we have chosen the natural unit $G=c=M=1$ in Appendix A and B.

A four-velocity can be assigned at any point in this space-time:
\begin{eqnarray} \label{eq:system2}
u^{t}&=&\frac{dt}{ds}=\frac{\exp^{-\nu}}{\sqrt{1-V^{2}}},
\nonumber \\
u^{\varphi}&=&\frac{d\varphi}{ds}=\varpi u^{t},
\nonumber \\
u^{r}&=&\frac{dr}{ds}=u^{t}v^{r},
\\
u^{\theta}&=&\frac{d\theta}{ds}=u^{t}v^{\theta},
\nonumber
\end{eqnarray}
where $\varpi=d\varphi/dt$, $v^{r}=dr/dt$ and $v^{\theta}=d\theta/dt$,
and
\begin{equation}
V^{2}={\exp}^{2\psi-2\nu}(\varpi-\omega)^{2}+{\exp}^{2\mu_{2}-2\nu}(v^{r})^{2}+
\exp^{2\mu_{3}-2\nu}(v^{\theta})^{2}.
\end{equation}
In the locally non-rotating inertial frame, the four velocity is given by
\begin{eqnarray} \label{eq:system3}
u^{(t)}&=&\gamma=\frac{1}{\sqrt{1-V^2}},
\nonumber \\
u^{(\varphi)}&=&\gamma\beta_{\varphi}=\frac{{\exp}^{\psi-\nu}(\varpi-\omega)}
{\sqrt{1-V^2}},
\nonumber \\
u^{(r)}&=&\gamma\beta_{r}=\frac{{\exp}^{\mu_2-\nu}v^r}{\sqrt{1-V^2}},
\\
u^{(\theta)}&=&\gamma\beta_{\theta}=\frac{{\exp}^{\mu_3-\nu}v^{\theta}}
{\sqrt{1-V^2}}.
\nonumber
\end{eqnarray}
Then the components of the comoving tetrad associated with this point are:
\begin{eqnarray} \label{eq:system4}
{\bf e}_{(\hat{t})}&=&{\bf u},
\nonumber \\
{\bf e}_{(\hat{\varphi})}&=&\gamma \beta_{\varphi}\exp^{-\nu}\partial_{\it t}+
[\gamma \omega \beta_{\varphi}\exp^{-\nu}+(1+\beta_{\varphi}^2\frac{\gamma-1}
{V^2})\exp^{-\psi}]\partial_{\varphi}+
\nonumber \\
& &\beta_{\varphi}\beta_{r}\frac{\gamma-1}{V^2}\exp^{-\mu_2}\partial_{\it r}+
\beta_{\varphi}\beta_{\theta}\frac{\gamma-1}{V^2}\exp^{-\mu_3}\partial_{\theta},
\nonumber \\
{\bf e}_{(\hat{r})}&=&\gamma \beta_{r}\exp^{-\nu}\partial_{\it t}+
(\gamma \omega \beta_{r}\exp^{-\nu}+\beta_{r}\beta_{\varphi}\frac{\gamma-1}
{V^2}\exp^{-\psi})\partial_{\varphi}+
\nonumber \\
 & & [1+\beta_{r}^2\frac{\gamma-1}{V^2}]\exp^{-\mu_2}\partial_{\it r}+
\beta_{r}\beta_{\theta}\frac{\gamma-1}{V^2}\exp^{-\mu_3}\partial_{\theta},
\\
{\bf e}_{(\hat{\theta})}&=&\gamma \beta_{\theta}\exp^{-\nu}\partial_{\it t}+
(\gamma \omega \beta_{\theta}\exp^{-\nu}+\beta_{\theta}\beta_{\varphi}
\frac{\gamma-1}{V^2}\exp^{-\psi})\partial_{\varphi}+
\nonumber \\
& & \beta_{\theta}\beta_{r}\frac{\gamma-1}{V^2}\exp^{-\mu_2}\partial_{\it r}+
[1+\beta_{\theta}^2\frac{\gamma-1}{V^2}]\exp^{-\mu_3}\partial_{\theta}.
\nonumber
\end{eqnarray}

The trajectory of any photon emitted from a flare with bulk motion
at this point can be specified by
two motion constants, i.e. the component of angular momentum parallel to
the symmetry axis $l$ and the Carter constant $Q$, which can be derived
from the direction of the photon momentum and the comoving tetrad:
\begin{eqnarray} \label{eq:system5}
\cos\Psi&=&-\frac{{\bf p}_{\rm s}\cdot {\bf e}_{(\hat{r})}}
{{\bf p}_{\rm s}\cdot{\bf u}_{\rm s}}, \nonumber \\
\sin\Psi\cos\Phi&=&-\frac{{\bf p}_{\rm s}\cdot {\bf e}_{(\hat{\theta})}}
{{\bf p}_{\rm s}\cdot{\bf u}_{\rm s}}, \\
\sin\Psi\sin\Phi&=&-\frac{{\bf p}_{\rm s}\cdot {\bf e}_{(\hat{\varphi})}}
{{\bf p}_{\rm s}\cdot {\bf u}_{\rm s}},
\nonumber
\end{eqnarray}
where ${\bf p}_{s}$ is the photon momentum at this location, and ${\bf u}_{s}$
is the bulk velocity of the flare at this point as expressed in (
\ref{eq:system3}). Based on equations (\ref{eq:system3}), (\ref{eq:system4})
and (\ref{eq:system5}), we can obtain
\begin{eqnarray} \label{eq:system6}
p_{\theta}&=&\frac{{\mathcal K}_1}{\mathcal D} l+\frac{{\mathcal B}_1}
{\mathcal D},\nonumber \\
p_{r}&=&\frac{{\mathcal K}_2}{\mathcal D} l+\frac{{\mathcal B}_2}
{\mathcal D},
\end{eqnarray}
where 
\begin{eqnarray} \label{eq:system7}
{\mathcal K}_1&=&\left[(1-\gamma)(f_2 \beta_{\theta}^2+f_1 f_2 
\beta_{r}\beta_{\theta}+f_1\beta_{\theta}\beta_{\varphi})+\gamma
f_2 V^2\right], \nonumber \\
& &\times\frac{\Sigma}{A^{1/2}\sin\theta}+\gamma \omega V^2(f_2
\beta_{\varphi}-f_1\beta_{\theta})\left(\frac{A}{\Delta}\right)^{1/2},
\nonumber \\
{\mathcal K}_2&=&\left[(1-\gamma)(f_1 f_2 \beta_{r}^2+f_1\beta_{r}
\beta_{\varphi}+f_2\beta_{r}\beta_{\theta})+\gamma f_1 f_2 V^2\right],
\nonumber \\
& &\times\frac{\Sigma}{(A\Delta)^{1/2}\sin\theta}+\gamma\omega f_1 V^2
(f_2 \beta_{\varphi}-\beta_{r})\frac{A^{1/2}}{\Delta},
\nonumber \\
{\mathcal B}_1&=&\gamma V^2 \left(f_1 \beta_{\theta}-f_2 \beta_
{\varphi}\right)\left(\frac{A}{\Delta}\right)^{1/2},
\nonumber \\
{\mathcal B}_2&=&\gamma f_1 V^2 \left(\beta_{r}-f_2 \beta_{\varphi}
\right)\left(\frac{A^{1/2}}{\Delta}\right),
\\
{\mathcal D}&=&
\gamma f_1 V^2+(1-\gamma)\left(f_1 \beta_{\varphi}^2+f_2
\beta_{\theta}\beta{\varphi}+f_1f_2\beta_{r}\beta_{\varphi}\right),
\nonumber \\
f_1&=&\frac{\cos\Psi}{\sin\Psi\cos\Phi},
\nonumber \\
f_2&=&\frac{\cos\Psi}{\sin\Psi\sin\Phi}. \nonumber
\end{eqnarray}

Combining the equations (\ref{eq:system6}-\ref{eq:system7})
and the photon momentum equations
\begin{eqnarray} \label{eq:system8}
\frac{dt}{d\lambda}&=&\frac{A}{\Delta\Sigma}(1-\omega l), \nonumber \\
\frac{dr}{d\lambda}&=&\pm \frac{1}{\Sigma}\left[(2r-al)^2+\Delta(r(r+2)-L)
\right]^{1/2}, \nonumber \\
\frac{d\theta}{d\lambda}&=&\pm \frac{1}{\Sigma}\left(L-\frac{l^2}{\sin^2\theta
}+a^2\cos^2\theta\right)^{1/2}, \\
\frac{d\varphi}{d\lambda}&=&\frac{A}{\Delta\Sigma}\left[\left(\frac{\Sigma
-2r}{A}\right)\frac{l}{\sin^2\theta}+\omega\right]^{1/2}, \nonumber 
\end{eqnarray}
we obtain the two motion constants $l$ and $Q$:
\begin{eqnarray} \label{eq:system9}
l&=&\frac{-Y+sgn(\pi-\Phi)\sqrt{Y^2-4XZ}}{2X},
\nonumber \\
Q&=&\frac{{\mathcal B}_1^2}{{\mathcal D}^2}+2\frac{{\mathcal B}_1{\mathcal K}_1}
{{\mathcal D}^2}l+(\frac{{\mathcal K}_1^2}{{\mathcal D}^2}+
\cot^2\theta)l^2-a^2\cos^2\theta,
\end{eqnarray}
where 
\begin{eqnarray} \label{eq:system10}
X&=&\frac{r^2-2r}{\Delta}+\Delta \frac{{\mathcal K}_2^2}{{\mathcal D}^2}+
\frac{{\mathcal K}_1^2}{{\mathcal D}^2}+\cot^2\theta,
\nonumber \\
Y&=&\frac{4ar}{\Delta}+2\Delta \frac{{\mathcal B}_2{\mathcal K}_2}
{{\mathcal D}^2}+2\frac{{\mathcal B}_1{\mathcal K}_1}{{\mathcal D}^2},
\\
Z&=&\Delta \frac{{\mathcal B}_2^2}{{\mathcal D}^2}+\frac{{\mathcal B}_1^2}
{{\mathcal D}^2}-\frac{r^4+a^2r^2+2ra^2}{\Delta}-a^2\cos^2\theta.
\nonumber 
\end{eqnarray}

The above expression (\ref{eq:system9}) for the two motion constants can be
applied to calculations of the photon trajectory of any point-like source with 
bulk motion in Kerr metric. If the source is static or has no bulk motion
(i.e. $\beta_{\varphi}\rightarrow 0$, $\beta_{r}\rightarrow 0$
and $\beta_{\theta}\rightarrow 0$), this expression reduces to
the simple one given by \citet[]{kvp92}.

For a flare above an accretion disk, where the flare bulk motion is observed
in the locally non-rotating frame to be predominantly in the $z$-direction
(say, upward or downward relative to
the accretion disk), then the $z$-direction velocity
is proportional to the $z$-direction unit vector:
\begin{equation}
{\bf u}_z\propto {\bf n}_z = {\exp}^{-\mu_{2}}\cos\theta \partial_{\it r}-
{\exp}^{-\mu_3}\sin\theta \partial_{\theta}.
\end{equation}
The four velocity of the source is given by:
\begin{equation}
{\bf u}=u^t({\partial_{\it t}}+\omega \partial_{\varphi} +V_{z}{\exp}^{\nu-\mu_2} \cos\theta
\partial_{\it r}-V_{z}{\exp}^{\nu-\mu_3}\sin\theta\partial_{\theta}),
\end{equation}
where $V_{z}$ is the $z$-direction velocity relative to the locally
non-rotating inertial frame. In the locally non-rotating inertial frame,
the velocity components are: 
\begin{eqnarray} \label{eq:system11}
\beta_{\varphi}&=&0,  
\nonumber \\
\beta_{r}&=&V_{z} \cos\theta,
 \\
\beta_{\theta}&=&-V_{z} \sin\theta.
\nonumber
\end{eqnarray}

If the X-ray flare motion is dominated by motion in the $r$-direction (say,
outward or inward), then the only non-zero velocity component in the locally
non-rotating inertial frame is $\beta_{r}=V_{r}$.
It is also possible that the flare co-rotates with the disk while other
direction motion can be neglected (i.e. $\beta_{r}=\beta_{\theta}=0$).
The motion constants in (\ref{eq:system9}) are then
equivalent to the ones given by \citet{rus00}.

Note that any choice of the value of $V_z$ or $V_{r}$ must meet the
requirement of $V^2<1$ in order that the source follows a time-like
world-line. 

For the special case that the flare is on axis and has only $r$-/$z$-direction
motion (i.e. $\theta \rightarrow 0$ and $\beta_{\varphi}=\beta_{\theta}=0$),
the two motion constants reduce to:
\begin{eqnarray} \label{eq:system12}
l&=&0,
\nonumber \\
Q&=&\frac{r^4+a^2r^2+2a^2r}{\Delta}-\frac{A}{\Delta}\left(\frac{\beta_{r}+
\cos\Psi}{\beta_{r}\cos\Psi+1}\right)^2.
\end{eqnarray}

\section{Illumination of the disk and the direct observed continuum flux}

We assume the X-ray luminosity in the source comoving frame to be isotropic
and in the form of a power law $L_{E_e}=L_0 E^{-\alpha}_{\rm e}$, where
$E_{\rm e}$ is the photon energy in the source comoving frame. 
The Monte-Carlo simulation shows that the anisotropic
effect is very weak for multiple scattering which dominates the hard X-ray
emission, although the soft X-ray emission is significantly anisotropic in the
case of single scattering approximation (Janiuk, Czerny \& \.{Z}ycki 2000).
In the frame rotating with the disk, including the `k-correction',
the illuminating flux on the disk element ($r\rightarrow r+dr$, $\varphi
\rightarrow \varphi+d\varphi$) with an incident angle of $\theta_{\rm in}$
($\cos\theta_{\rm in}=-{\bf p}\cdot{\bf n}/{\bf p}\cdot{\bf u_{\rm d}}$,
where ${\bf n}$ is the equator normal)
is given by (e.g. Yu \& Lu 2000; Ruszkowski 2000):
\begin{equation}
F^{\rm ill}_{\rm d}(E_{\rm d}, \theta_{\rm in}; r,\varphi) = L_0 g^{1+\alpha}_{
\rm sd} E^{-\alpha}_{\rm d} \gamma^{-1} \frac{f_{\rm sd}}{d\varphi dr}(
\frac{\Delta}{A})^{1/2},
\end{equation}
where $f_{\rm sd}$ is the fraction of total photons which corresponds to
the grid elements defined by $d\varphi dr$, $g_{\rm sd}$ is the redshift factor
for a photon propagating from the source to the disk grid elements given by
\begin{equation}
g_{\rm sd}=\frac{{\bf p}_{\rm s}\cdot {\bf u}_{\rm s}}{{\bf p}_{\rm d}
\cdot {\bf u}_{\rm d}},
\end{equation}
and $\gamma$ is the Lorentz factor of the relative motion of the disk element
and the locally non-rotating inertial frame. For $r\geq r_{\rm ms}$, where
$r_{\rm ms}$ is the radius of the last stable orbit, the $\gamma$ factor is
given by:
\begin{equation}
\gamma=(1-V^{(\varphi)2})^{-1/2}, \ \ \ V^{(\varphi)}=\frac{A}{(
\Sigma \Delta)^{1/2}}(\Omega_d-\omega),
\end{equation}
where $\Omega_{\rm d}=1/(a+r^{3/2})$ is the rotation velocity of the disk
(see Cunningham 1975). The incident photon number intensity is then
$N_{E_{\rm d}}^{\rm in}(E_{\rm d},\theta_{\rm in};r,\varphi)= F^{\rm ill}_{\rm d}(E_{\rm d}, \theta_{\rm in}; r,\varphi)/E_{\rm d}$.

Note in equation (B1) the power index of $g$ is $(1+\alpha)$, which is 
different from $(2+\alpha)$ in \citet[]{yl00} where the total 
flux is considered.

If a distant observer is located at $(r_{\rm o},\theta_{\rm o},\varphi_{\rm o})$ with collecting
area on the $r=r_{\rm o}$ sphere equal to
\begin{equation}
\delta S=r^2_{\rm o}\sin \theta_{\rm o} \delta \varphi_{\rm o} \delta \theta_{\rm o},
\end{equation}
the direct flux component from the flare received by the observer is
analogously given by
\begin{equation}
F_{\rm dir}(E_{\rm obs})= L_0 g^{1+\alpha}_{
\rm so} E^{-\alpha}_{\rm o}\frac{f_{\rm so}}{\delta S},
\end{equation}
where $f_{\rm so}$ is the fraction of the total number of emitted photons 
intersecting the observers collecting area, and $g_{\rm so}$ is the redshift
factor of a photon propagating from the flare to the observer and is given
by a formula similar to equation (B2):
\begin{equation}
g_{\rm so}=\frac{{\bf p}_{\rm s}\cdot {\bf u}_{\rm s}}{{\bf p}_{\rm o}
\cdot {\bf u}_{\rm o}}.
\end{equation}


\end{document}